\documentclass[prb,twocolumn,amsmath,amssymb,superscriptaddress]{revtex4}
\usepackage[dvips]{graphicx}
\usepackage{wasysym}

\begin{document}

\title{Pyrochlore Photons: The U(1) Spin Liquid in a $S=1/2$ Three-Dimensional Frustrated Magnet}
\author{Michael Hermele}
\affiliation{Department of Physics, University of California, Santa Barbara, CA
 93106-9530}
\author{Matthew P. A. Fisher}
\affiliation{Kavli Institute for Theoretical Physics, University of 
California, Santa Barbara, CA 93106-4030}
\author{Leon Balents}
\affiliation{Department of Physics, University of California, Santa Barbara, CA
 93106-9530}
\date{\today}

\begin{abstract}
We study the $S=1/2$ Heisenberg antiferromagnet on the pyrochlore lattice in the limit of
strong easy-axis exchange anisotropy. We find, using only standard techniques of degenerate
perturbation theory, that the model has a $U(1)$ gauge symmetry generated by certain local
rotations about the $z$-axis in spin space. Upon addition of an extra local interaction in this
and a related model with spins on a three-dimensional
network of corner-sharing octahedra, we can write
down the exact ground state wavefunction with no further approximations. Using the properties of
the soluble point we show that these models enter the $U(1)$ spin liquid phase, a novel
fractionalized spin liquid with an emergent $U(1)$ gauge structure. This phase supports
gapped $S^z = 1/2$ spinons carrying the $U(1)$ ``electric'' gauge charge,
a gapped topological point defect or ``magnetic'' monopole, and a gapless
``photon,'' which in spin language is a gapless, linearly dispersing $S^z = 0$ collective mode.
There are power-law spin correlations with a nontrivial angular dependence, as well as novel
$U(1)$ topological order.
This state is stable to \emph{all} zero-temperature
perturbations and exists over a finite extent of
the phase diagram. Using a convenient lattice version of electric-magnetic duality, we develop
the effective description of the $U(1)$ spin liquid and the adjacent soluble point in terms of
Gaussian quantum electrodynamics and calculate a few of the universal properties. The resulting
picture is confirmed by our numerical analysis of the soluble point wavefunction. Finally, we
briefly discuss the prospects for understanding this physics in a wider range of models and 
for making contact with experiments.
\end{abstract}
\maketitle

\section{Introduction}
\label{sec:intro}

The search for quantum spin liquid states in frustrated magnets can be 
traced back at least as far as the early suggestion of a resonating 
valence bond state in the triangular lattice Heisenberg model\cite{pwa1}. 
Almost 15 years later, the suggestion of such a state in the undoped 
high-$T_c$ cuprates\cite{pwa2} set off an explosion of interest in 
two-dimensional spin liquids (i.e.~Mott insulators at half-filling 
with \emph{no} broken symmetries). Frustrated Heisenberg models on the square, 
triangular and kagom\'{e} lattices have all received significant 
attention as candidate systems for quantum disordered ground states. 
While there has been comparatively little theoretical work on quantum spin liquids
in three-dimensional frustrated magnets, materials with magnetic ions on the pyrochlore 
lattice (Fig.~\ref{fig:pyro}) may
be good candidates for spin liquids and other exotic states. To give one example,
recent neutron scattering experiments
on ${\rm ZnCr_2O_4}$, a $S = 3/2$ pyrochlore
Heisenberg antiferromagnet, suggest a nontrivial disordered state\cite{zinc-chromate1}
above a transition to N\'{e}el order accompanied by a lattice distortion\cite{zinc-chromate2}
at 12.5 K.

Meanwhile, much work has been devoted to understanding the properties 
of possible spin liquid states, independent of their existence in 
particular microscopic models. Most of the proposed 
spin liquid states support deconfined $S = 1/2$ spinons; such states 
are fractionalized in that some of the elementary excitations carry 
quantum numbers that are fractions of those allowed in a finite-size 
system. Fractionalized states can be precisely characterized by their 
\emph{topological order}\cite{wen-niu}, which in the simplest scenario is associated 
with the topological sectors of an emergent, deconfining $Z_2$ gauge 
field\cite{sf-topord}. 
The ``vison,'' a gapped vortex-like excitation that carries the $Z_2$ 
flux\cite{read-chakraborty}, must also be present.
While topological order does not require a 
liquid ground state and can coexist with conventional long-range order, 
we believe it is probably common in spin liquids and hence in some of the 
nearby ordered states. Very recently, many of these ideas have been 
put on firmer ground by the emergence of several microscopic models 
supporting stable fractionalized phases in two and three dimensions
\cite{tri-rvb,kag,lesik-senthil1,lesik-senthil2,artificial-light,lesik-z3,
kagome-dimer-model}.

Despite these recent theoretical successes, an unambiguous experimental 
realization of these ideas is still lacking. Indeed, spin liquid states 
seem rather rare; is topological order rare as well? For 
$Z_2$-fractionalized states this question is difficult to answer, because 
the gapped visons have no effect on easily measurable low-energy 
properties. Clever 
proposals have been made\cite{sf-senthil-expt} and carried out\cite{visonexp1,visonexp2}
to directly detect topological order 
in the cuprates (with negative results thus far), but these experiments 
are difficult and rely on properties of the phases proximate to a 
topologically ordered state. $Z_2$ topological order is difficult enough 
to observe that it is impossible to say at present how rare or common it 
is.

Fortunately it may be possible to shed new light on the
experimental situation. In this paper, we present two models of
\emph{three-dimensional} $S = 1/2$ frustrated magnets, one on the 
pyrochlore lattice, the other on a related network of corner-sharing 
octahedra (the links of the cubic lattice,
as shown in Fig.~\ref{fig:cubic-octahedra}).
Both of these models exhibit a novel fractionalized phase, the 
$U(1)$ spin liquid. This state has an emergent $U(1)$ gauge 
structure that gives rise to several remarkable properties: there is a 
\emph{gapless} 
``artificial photon'' excitation, a gapped spinon carrying ``electric'' 
gauge charge, a gapped ``magnetic'' monopole, an emergent $1/r$ 
``Coulomb'' potential between pairs of spinons and monopoles, and novel 
$U(1)$ topological order.
If this phase exists in a real material, the gapless photon should have 
important implications for low-energy thermodynamics, transport 
and spectroscopy; therefore $U(1)$ fractionalization may be easier to find 
in experiments. Such states, thus far realized in 
large-$N$ spin models\cite{origin-of-light}
and bosonic Hubbard-type models\cite{lesik-senthil1,lesik-senthil2,artificial-light},
arise as the 
deconfined or Coulomb phase of compact $U(1)$ lattice gauge 
theory. While most work on spin liquids has focused on
$d \leq 2$, motivated by the 
cuprates and the conventional wisdom that quantum fluctuations are more 
effective at destroying long range order in low dimensions, the $U(1)$ 
spin liquid \emph{only} occurs in $d \geq 3$; for $d \leq 2$ 
the Coulomb phase of compact $U(1)$ gauge theory (with gapped matter)
is always unstable due to instanton effects\cite{polyakov}.

\begin{figure}
\begin{center}
\centerline{\includegraphics[width=3.3in]{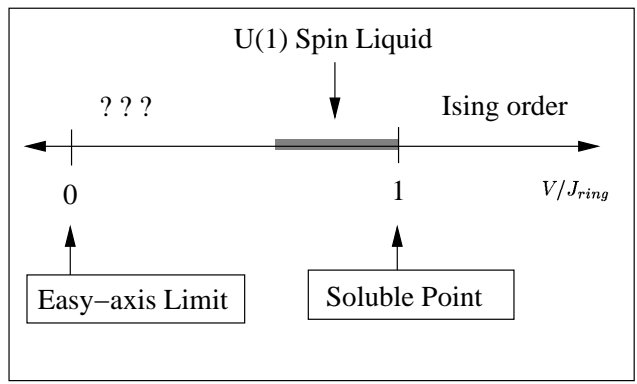}}
\caption{Phase diagram for both models. The parameter $V/J_{ring}$
is the relative 
strength of the Rokhsar-Kivelson potential and the XY ring exchange that obtains in the
easy-axis limit of the Heisenberg model. The soluble point is located at $V/J_{ring} = 1$,
which is a special deconfined point of the adjacent $U(1)$ spin liquid. Just to the right of
the soluble
point the models go into an Ising ordered state. Sufficiently far to the left we expect
Ising order, while at intermediate values of $V/J_{ring}$ states with broken
translation symmetry but no magnetic order are also possible.
Immediately to the left of the soluble point,
the $U(1)$ spin liquid exists over a finite (but unknown) extent of the phase diagram.}
\label{fig:phasediag}
\end{center}
\end{figure}

Both models are of intrinsic interest as examples of tractable but 
nontrivial frustrated magnets. The pyrochlore model is particularly 
appealing due to its simplicity:  its derivation begins with the 
nearest-neighbor $S = 1/2$ Heisenberg antiferromagnet. Taking the limit of 
large easy-axis exchange anisotropy $J_z \gg J_\perp$ simplifies the 
problem by breaking the spectrum into extensively degenerate manifolds 
with large separations of $O(J_z)$. It is then possible 
to write an effective Hamiltonian describing the splitting of the low-energy manifold, 
using standard techniques of degenerate perturbation theory in $J_\perp$. 
This effective Hamiltonian \emph{has} a $U(1)$ gauge structure, which 
forms the foundation for our subsequent analysis\cite{roderich-pc}.
Another point of view,
equivalent at the level of perturbation theory but perhaps with broader
implications in more general scenarios, is that the low-energy
sector of the model is \emph{unitarily equivalent} to a $U(1)$ gauge theory.
It is not obvious how 
to treat the resulting model analytically, but upon addition of an extra 
six-site interaction term it can be tuned to a soluble point where 
it is possible to write an exact ground state wavefunction with no further 
approximations. The models can be reinterpreted as quantum dimer models (QDMs), and the
extra term as the analog of the Rokhsar-Kivelson (RK) potential in the square lattice 
QDM\cite{rk}.
As will be explained in detail below, the properties 
of the soluble point allow us to locate the $U(1)$ spin liquid adjacent to it. 
Since this state is stable to \emph{all} zero-temperature
perturbations, it persists over 
a finite extent of the phase diagram (Fig.~\ref{fig:phasediag}). Furthermore, stability to large
but finite $J_z$ implies that the $U(1)$ gauge structure persists in the absence of microscopic
local symmetries and is truly emergent. 
On the purely theoretical side, we 
believe these models give the first examples of $U(1)$ gauge theories that 
have a deconfining phase even in the limit of infinitely strong 
bare coupling. The first such $Z_2$ gauge theory was 
discovered only recently by Moessner and Sondhi\cite{tri-rvb}.

The effective theory of the $U(1)$ spin liquid and the soluble RK point
is simply Gaussian quantum electrodynamics (QED). At the RK
point, which is itself a special deconfined limit of the generic phase,
the ``electric stiffness,'' or coefficient of ${\bf E}^2$ in the Hamiltonian, vanishes.
This is a higher-dimensional generalization of the
effective picture of the square lattice QDM in terms of a coarse-grained height field
\cite{henley-unpub}.

The $U(1)$ spin liquid has power-law correlations with nontrivial angular dependence,
novel $U(1)$ topological order,
and supports gapped $S^z = 1/2$ spinons, a gapped topological point
defect (the ``magnetic'' monopole) and a gapless $S^z = 0$ collective mode corresponding to
the photon of the gauge theory.
The latter excitation makes an additive $T^3$ contribution
to the low-temperature specific heat, and should affect various other low-energy properties
of $U(1)$-fractionalized phases (either the $U(1)$ spin liquid, or phases with coexisting
conventional and topological order). If such a phase exists in a real material, we speculate
that it may be possible to probe ``photons'' with photons via Raman scattering.

\subsection{Outline}

We begin Sec.~\ref{sec:models} with a derivation of the pyrochlore model
starting from the Heisenberg antiferromagnet. In Sec.~\ref{sec:cubic-model}
the cubic (or corner-sharing octahedra) model is discussed. The remainder of Sec.~\ref{sec:models}
is concerned with demonstrating the equivalence of the spin models to frustrated compact $U(1)$
gauge theories, and developing a useful lattice version of electric-magnetic duality.

Beginning from the dual description, Sec.~\ref{sec:effective-theory} develops the 
effective description of the $U(1)$ spin liquid
and the soluble point in terms of Gaussian quantum electrodynamics. Corrections to effective
action and to the scaling equalities between microscopic and effective degrees of freedom are
discussed in Sec. \ref{sec:corrections}. Sec.~\ref{sec:properties} contains a
discussion of the universal properties of the $U(1)$ spin liquid, including its novel
$U(1)$ topological order. In 
Sec.~\ref{sec:rkpoint} we present our analysis of the soluble point ground state wavefunction,
which gives strong support for the validity of our effective picture. We conclude in 
Sec.~\ref{sec:discussion} with a discussion of open issues, focusing on the challenging problems
of understanding this physics in a broader range of models and looking for $U(1)$-fractionalized
phases in real materials.

\section{Models and Mappings}
\label{sec:models}

\subsection{Pyrochlore Model}
\label{sec:pyro-model}

\begin{figure}
\begin{center}
\centerline{\includegraphics[width=3.3in]{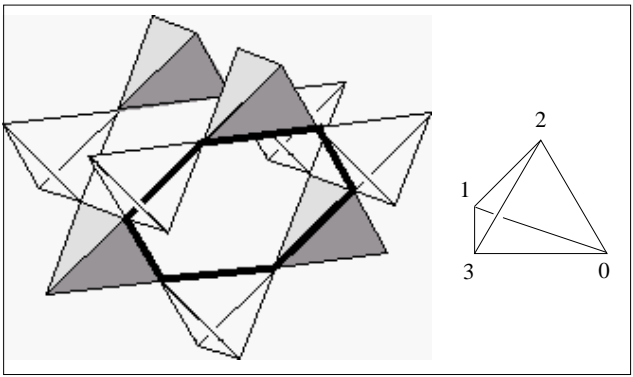}}
\caption{The pyrochlore lattice (left), and one up-pointing tetrahedron (right).
One sublattice of tetrahedra is shaded, and the other transparent. The thickened
bonds show the location of a pyrochlore hexagon. Each such hexagon is a member
of one of four orientations of kagom\'{e} lattice planes. The numbering
of sites in the up-pointing tetrahedron on the right is the convention used in the
text. For $i=0,1,2$, the fcc Bravais lattice vector ${\bf a}_i$ points in the
direction given by looking from site $3$ to site $i$.}
\label{fig:pyro}
\end{center}
\end{figure}

We begin with the nearest-neighbor $S = 1/2$ Heisenberg antiferromagnet on 
the pyrochlore lattice. This structure is a three-dimensional network of 
corner-sharing tetrahedra (Fig.~\ref{fig:pyro}). It can be obtained by translating one
``up-pointing'' tetrahedron (shown on the right of Fig.~\ref{fig:pyro})
through the fcc Bravais lattice vectors ${\bf R} = n_0 {\bf a}_0 + 
n_1{\bf a}_1 + n_2{\bf a}_2$. We choose ${\bf a}_0 = {\bf x}$, ${\bf 
a}_1 = {\bf x}/2 + \sqrt{3}{\bf y}/2$, and ${\bf a}_2 = {\bf x}/2 + 
{\bf y}/2\sqrt{3} + \sqrt{2/3}{\bf z}$. Basis vectors for the reciprocal 
lattice are defined by ${\bf b}_i = \sqrt{2}\pi \epsilon_{ijk} {\bf a}_j 
\times {\bf a}_k$, so that ${\bf a}_i \cdot {\bf b}_j = 2\pi \delta_{ij}$. 
The four sites in each unit cell are distinguished by an index $i
= 0,\ldots,3$, as indicated in Fig.~\ref{fig:pyro}. Lattice sites are denoted
either by single italic letters like $i$, or by pairs $({\bf R},i)$ when we wish to
specify the position of a site within the unit cell.

Up to a constant 
the Hamiltonian can be written as a sum over tetrahedra:
\begin{equation}
\label{eq:pyro-su2}
{\cal H} = \frac{J}{2} \sum_t ({\bf S}_t)^2 \text{,}
\end{equation}
where ${\bf S}_t = \sum_{i \in t} {\bf S}_i$ is the total spin on the
tetrahedron $t$. Following the analysis of 
a generalized kagom\'{e} Heisenberg antiferromagnet in 
Ref.~[\onlinecite{kag}], we introduce easy-axis exchange anisotropy:
\begin{eqnarray}
\label{eq:pyro-easy-axis}
{\cal H} &=& {\cal H}_I + {\cal H}', \\
{\cal H}_I &=& \frac{J_z}{2} \sum_t (S^z_t)^2, \\ 
{\cal H}' &=& \frac{J_\perp}{2} \sum_{\langle i j \rangle} (S^+_i S^-_j + h.c.),
\end{eqnarray}
where $J_z \gg J_\perp$. This reduces the global $SU(2)$ invariance to 
$U(1) \times Z_2$. We first consider the point $J_\perp = 0$,
where ${\cal 
H}$ reduces to a classical Ising model, with ground states specified by 
$S^z_t = 0$ on all tetrahedra. It was argued by Anderson\cite{pwa3} that,
almost 
identically to Pauling's model for water ice\cite{pauling}, this Ising 
model has an extensive ground state degeneracy (i.e.~finite $T=0$ entropy 
per site).

\begin{figure}
\begin{center}
\centerline{\includegraphics[width=3.3in]{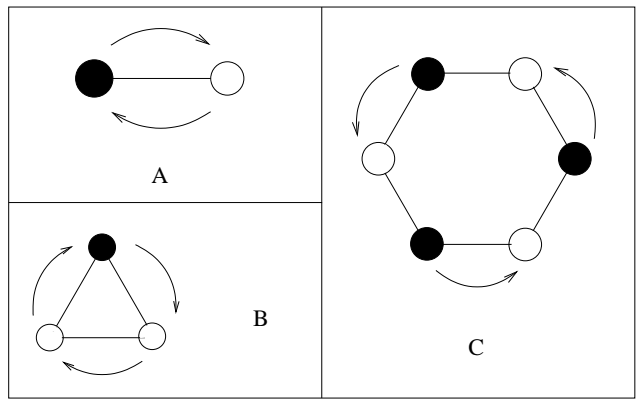}}
\caption{Depiction of the processes contributing to the third-order degenerate
perturbation theory for the easy-axis pyrochlore Heisenberg antiferromagnet. Processes
(A) and (B) give only trivial constant shifts of the energy. Process (C) leads to an XY ring
exchange term acting on hexagonal plaquettes.}
\label{fig:pyro-pt}
\end{center}
\end{figure}

A small $J_\perp > 0$ introduces quantum fluctuations and lifts the 
extensive degeneracy; this splitting is encapsulated in an effective 
Hamiltonian using standard techniques of perturbation theory.
The first-order contribution is easily seen to vanish. We will need to go 
to third order, where we have the general expression:
\begin{equation}
\label{eq:generalpt}
{\cal H}_{eff} = (1 - {\cal P}){\Big [}
- {\cal H}' \frac{{\cal P}}{{\cal H}_I} {\cal H}' 
+ {\cal H}' \frac{{\cal P}}{{\cal H}_I} {\cal H}' \frac{{\cal P}}{{\cal H}_I} 
  {\cal H}'
{\Big ]}(1 - {\cal P}) \text{.}
\end{equation}
Here ${\cal P}$ projects onto the orthogonal complement of the ground 
state manifold. To describe the processes contributing in 
Eq.~(\ref{eq:generalpt}), it is useful to work in the standard hardcore 
boson language for the spins, where $S^z = \pm 1/2$ corresponds to the 
presence/absence of a boson. Each term in ${\cal H}'$ hops bosons along 
nearest-neighbor bonds; acting on a state in the low-energy manifold, each 
hop creates two tetrahedra with $S^z_t \neq 0$. At second order in
${\cal H}'$, bosons 
can hop and then return along the same bond (Fig.~\ref{fig:pyro-pt}A). 
This can always occur on 4 bonds in every tetrahedron, 
thus giving only a constant contribution to the energy. At third order 
another constant contribution arises from single bosons (or holes) hopping 
around triangular faces (Fig.~\ref{fig:pyro-pt}B). There is also a 
nontrivial \emph{ring exchange} process acting on the hexagonal plaquettes 
(see Fig.~\ref{fig:pyro}), where hexagons containing three evenly spaced 
bosons can be rotated as shown in Figure~\ref{fig:pyro-pt}C. The resulting 
effective Hamiltonian is
\begin{eqnarray}
\label{eq:pyro-heff-with-constants}
{\cal H}_{eff} &=& (J_\perp^2/J_z)(J_\perp/J_z - 1) N_t
 \\
&+& J_{ring} \sum_{\hexagon} (S_1^+ S_2^- S_3^+ S_4^- S_5^+ 
S_6^- + 
h.c.) \nonumber \text{,}
\end{eqnarray}
where $N_t$ is the total number of tetrahedra,
$J_{ring} = 3 J_\perp^3/2J_z^2$, and the sum is over hexagonal 
plaquettes. The labelling of the spin operators inside the sum is given by 
moving around each hexagon in an arbitrary direction.
Note that $[{\cal H}_{eff}, S^z_t] = 0$, as must 
be true for \emph{any} effective Hamiltonian acting in the low-energy 
manifold, whatever the form of ${\cal H}'$.

We focus on the extreme easy axis limit described by ${\cal H}_{eff}$, but note 
in passing that a finite but large $J_z$ would introduce small
fluctuations out of the ground state manifold. While these will not affect
universal properties, they can matter for the short-distance correlation
functions of some microscopic operators. This can be understood formally by
a more sophisticated execution of the perturbation theory in $J_{\perp}$
that accounts for splitting of the low-energy manifold and mixing of higher states
on an equal footing\cite{mike-unpub}.
The main result is that the problem at finite $J_z$ can be mapped, by a unitary
transformation, order-by-order in $J_{\perp}$ onto a transformed Hamiltonian acting only
within the low-energy manifold where $S^z_t = 0$. 
This mapping accounts for finite $J_z$ by generating nontrivial
perturbative relations between physical and transformed operators; however, we
ignore these corrections for simplicity and use only the results of the standard degenerate
perturbation theory described above.

It is possible, and will be convenient, to change the sign of the ring
term by a similarity transformation. On any given 
site we can make the transformation $S^z \rightarrow S^z$ and $S^\pm 
\rightarrow - S^\pm$ by making a $\pi$-rotation about the $z$-axis in 
spin space. One transformation with the desired effect, consisting of 
$\pi$-rotations on a pattern of sites, is:
\begin{eqnarray}
\label{eq:pyro-similarity-transformation}
S^z_i &\rightarrow& S^z_i \\
S^\pm_{{\bf R}i} &\rightarrow&
\operatorname{exp}(i {\bf Q}_i \cdot {\bf R})
S^\pm_{{\bf R}i}\text{,}
\end{eqnarray}
where ${\bf Q}_0 = {\bf Q}_1 = ({\bf b}_1 + {\bf b}_2)/2$ and
${\bf Q}_2 = {\bf Q}_3 = 0$.

After this transformation the Hamiltonian takes the form
\begin{equation}
\label{eq:pyro-heff}
{\cal H}_p = - J_{ring} \sum_{\hexagon} (S_1^+ S_2^- S_3^+ S_4^- 
S_5^+ S_6^- + h.c.)\text{,}
\end{equation}
where the constant terms have been dropped. Models similar to this one on 
the kagom\'{e}\cite{kag}, square\cite{ebl,anders}, triangular\cite{triring} 
other lattices\cite{lesik-senthil2},
where XY ring exchange of spins or bosons is a dominant term, 
have recently been shown to exhibit a variety of unusual phases and 
critical behavior. The physics of the pyrochlore ring exchange model 
should be accessible to quantum Monte Carlo studies; while the original 
Hamiltonian in Eq.~(\ref{eq:pyro-easy-axis}) has a sign problem, ${\cal 
H}_p$ does not.

\begin{figure}
\begin{center}
\centerline{\includegraphics[width=3.3in]{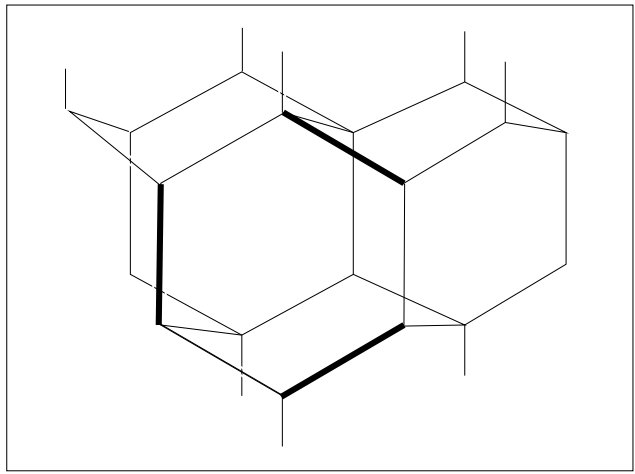}}
\caption{A small piece of the diamond lattice. The links form hexagonal loops 
corresponding to the pyrochlore hexagons. These are the shortest possible closed 
paths on the diamond lattice. The hexagon with three thickened bonds depicts
the dimer positions on a flippable hexagon. The alternating full and empty bonds correspond
to alternating up and down spins.}
\label{fig:diamond}
\end{center}
\end{figure}

${\cal H}_p$ can be reinterpreted as a quantum dimer model on the 
diamond lattice (Fig.~\ref{fig:diamond}), with two dimers touching every site. To see 
this, observe that the centers of the pyrochlore tetrahedra form a diamond 
lattice. Each nearest-neighbor diamond link passes through exactly one 
pyrochlore site, so we can reinterpret the pyrochlore spins as diamond 
link variables. The smallest closed loops in this lattice contain six 
links and correspond to the pyrochlore hexagons. We say a dimer is present 
on a given bond if $S^z_i = 1/2$, 
or absent if $S^z_i = -1/2$. $S^z_t = 0$ becomes the constraint that 
every diamond \emph{site} touches two dimers, and the ring exchange move 
is the most local dynamics preserving this constraint. Each term 
in ${\cal H}_p$ acts on a
``flippable'' hexagon, one containing alternating full and empty bonds 
as in Fig.~\ref{fig:diamond}, by rotating the dimers 
around it. Non-flippable hexagons are annihilated. 

As first realized by Rokhsar and Kivelson, dimer models generically have a 
point in their parameter space where it is possible to write down the 
exact ground state wavefunction\cite{rk}. To reach this point in our 
model, we add the term ${\cal H}_V = V N_f$, where 
$N_f$ is the number of flippable hexagons. The RK
point obtains for ${\cal H}_{RK} = {\cal H}_p + J_{ring}N_f$
(i.e. $V = J_{ring}$), and the ground state is an 
equal-weight superposition of all possible dimer coverings of the lattice 
that satisfy the constraint of two dimers touching every site. In the spin 
language, this wavefunction can be written as the projection of a 
transverse ferromagnet:
\begin{equation}
\label{eq:rk-wavefunction}
|\psi_{RK}\rangle = (1 - {\cal P}) \prod_i | S^x_i = 1/2\rangle \text{,}
\end{equation}
where, as in Eq.~(\ref{eq:generalpt}), $(1 - {\cal P})$ projects onto the 
$S^z_t = 0$ manifold. For completeness, we also express $N_f$ in 
terms of spin operators:
$N_f = \sum_{\hexagon} P_{flip}(\hexagon)$,
where $P_{flip}(\hexagon)$ gives unity acting on a flippable 
hexagon and zero otherwise. One has:
\begin{equation}
\label{eq:pyro-pflip}
P_{flip}(\hexagon) = \sum_{\sigma = \pm 1}
\prod_{j \in \hexagon} \big(\frac{1}{2} + \sigma (-1)^j S^z_j \big) 
\text{.}
\end{equation}
We will be interested in the properties of the generalized ring model 
${\cal H}_p + {\cal H}_V$ in the vicinity of the soluble point.

\subsection{Cubic Model}
\label{sec:cubic-model}

\begin{figure}
\begin{center}
\centerline{\includegraphics[width=3.3in]{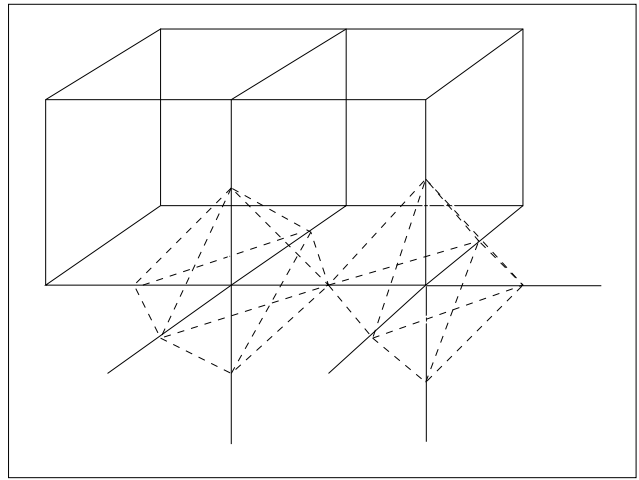}}
\caption{Illustration that the \emph{links} of the cubic lattice
are equivalent to the \emph{sites} of a lattice of corner-sharing octahedra.}
\label{fig:cubic-octahedra}
\end{center}
\end{figure}

Largely to simplify the geometry of the presentation, we introduce an 
alternate model that we find has many of the same properties as its 
pyrochlore analog. The model is the QDM on the cubic lattice, with 
\emph{three} dimers touching every site. We consider only the most local 
dynamics, which rotates the configuration on square plaquettes with two 
dimers on opposite sides, and the corresponding Rokhsar-Kivelson potential 
that counts flippable squares. Reversing the mapping above, we can also 
think of this as a spin model with $S = 1/2$ on the links of a cubic 
lattice, or, equivalently, on the sites of a lattice of corner-sharing 
octahedra with their centers at the cubic sites (Fig.~\ref{fig:cubic-octahedra}). The 
octahedra play the role of the pyrochlore tetrahedra, with the total spin 
on each $S^z_{oct} = 0$. We denote cubic sites by 
boldface letters like ${\bf r}$ and identify the links by specifying either pairs of adjacent
sites, or one site and the direction of the link. For example, the link connecting
a site ${\bf r}$ with its nearest neighbor in the positive $x$-direction (${\bf r}+{\bf x}$)
is denoted by $({\bf r},{\bf r}+{\bf x})$ or $({\bf r},x)$. Using this notation, we 
express the dimer kinetic term as a 4-site XY ring exchange for the spins:
\begin{eqnarray}
\label{eq:cubic-ring-model}
{\cal H}_c &=& -J_{ring} \sum_{\square}(S^+_1 S^-_2 S^+_3 S^-_4 + h.c.) \\
&=& -J_{ring} \sum_{{\bf r}} \Big( S^+_{{\bf r}x} S^-_{{\bf 
r}+{\bf x},y} S^+_{{\bf r}+{\bf y},x} S^-_{{\bf r}y} + h.c. + \cdots \nonumber
\Big) \text{,}
\end{eqnarray}
where the numbering in the first line runs around the perimeter of each square
plaquette, and in the second line only one orientation of plaquette is shown explicitly.
As before, we will be interested in the vicinity of the soluble point of ${\cal H}_c + 
{\cal H}_V$.

It is also interesting to note that, as for ${\cal H}_p$, the pure ring 
exchange model ${\cal H}_c$ can be derived as the easy-axis limit of a 
Heisenberg antiferromagnet. In this case one begins with spins on the 
network of corner-sharing octahedra with nearest-neighbor exchange, and an 
additional exchange of the same sign and magnitude between spins at 
opposite points of the same octahedron. The Hamiltonian can be written as 
a sum over octahedra and the analysis proceeds exactly as before.

\subsection{Frustrated $U(1)$ Gauge Theory}
\label{sec:frust-gt}

Both the cubic and pyrochlore models have an exact $U(1)$ gauge 
invariance, as is generally the case in dimer models. In the pyrochlore 
model, this arises because of the local integer conserved quantity 
$S^z_t$. 
The local symmetry is generated by rotations about the $z$-axis in spin 
space on all the sites in a given tetrahedron: $G_t(\alpha) = 
\operatorname{exp}(i \alpha S^z_t)$. These generators commute with 
one another and the Hamiltonian. It is important to emphasize that this 
gauge symmetry is generated by \emph{physical} transformations and has 
nothing to do with any redundancy in our description. These statements 
also hold for the cubic model, with octahedra substituted for 
tetrahedra. For now, we shall focus on the cubic model for ease of 
presentation. All of the statements in this and the following section can 
be generalized simply to the pyrochlore case; this is summarized in 
Section \ref{sec:pyro-mappings} below.

The gauge structure suggests we may gain insight by 
thinking about the model as a $U(1)$ lattice gauge theory, an 
approach that proved helpful in understanding the square lattice QDM 
\cite{fradkin-kivelson}. In fact, it is possible to formally rewrite ${\cal H}_c$ as 
a pure gauge 
theory on the cubic lattice. To see this, we find it convenient to 
soften the hardcore constraint on the bosons by explicitly introducing an 
on-site   
repulsion term in the Hamiltonian. We also go to quantum rotor variables
$n_{{\bf r}{\bf r}'} \in {\mathbb Z}$
and $\phi_{{\bf r}{\bf r}'} \in [-\pi,\pi)$, with commutation relations 
$[\phi_{{\bf r}i},n_{{\bf r}'j}] = i \delta_{ij} \delta_{{\bf r}{\bf r}'}$. (Here
and elsewhere we commit a standard abuse of notation and formally denote a constraint on
the eigenvalues of an operator as a constraint on the operator itself.)
Using the 
relations $S^z = n - 1/2$ and $S^\pm = \operatorname{exp}(\pm i \phi)$, we have
\begin{eqnarray}
\label{eq:cubic-soft}
{\cal H}_c &=& \frac{U}{2} \sum_{\langle{\bf r}{\bf r}'\rangle}
(n_{{\bf r}{\bf r}'} - 1/2)^2 \\
 &-& K \sum_{\square}
\operatorname{cos}(\phi_1 - \phi_2 + \phi_3 - \phi_4)\nonumber\text{,}
\end{eqnarray}
where the numbering inside the cosine proceeds around the perimeter of the 
given square. This is a faithful representation of the 
spin model Eq.~(\ref{eq:cubic-ring-model}) in the limit $U/K \rightarrow 
\infty$, which just imposes the hardcore constraint $n_{{\bf r}{\bf r}'} = 0,1$.

We now define an orientation on the cubic links, which we take to point 
out of the A sublattice and into the B sublattice. This allows us to define 
oriented link variables by:
\begin{eqnarray}
\label{eq:vector-vars-e}
e_{{\bf r}{\bf r}'} &=& \pm (n_{{\bf r}{\bf r}'} - 1/2) \\
\label{eq:vector-vars-a}
a_{{\bf r}{\bf r}'} &=& \pm \phi_{{\bf r}{\bf r}'} \text{.}
\end{eqnarray}
Here we take the plus/minus sign when ${\bf r}$ lies in the A/B 
sublattice.
Since $e_{{\bf r}{\bf r}'} = - e_{{\bf r}'{\bf r}}$, these variables can 
be thought of as components of vector fields taken along the links of the 
lattice.
Putting these definitions into the Hamiltonian, we have
\begin{equation}
\label{eq:cubic-gt}
{\cal H}_c = \frac{U}{2} \sum_{\langle{\bf r}{\bf r}'\rangle} 
e^2_{{\bf r}{\bf r}'}
- K \sum_{\square} 
\operatorname{cos}\Big(
\sum_{{\bf r}{\bf r}'\in\square}\!\!\!\!\!\!\!\!\!\!\!\circlearrowleft
\; a_{{\bf r}{\bf r'}} \Big) \text{.}
\end{equation}
The sum inside the cosine is taken in an oriented fashion around the links 
of the given square plaquette, and is thus a discrete line integral of 
$a_{{\bf r}{\bf r}'}$ along a smallest possible closed path. This is a lattice
version of the curl, so we define:
\begin{equation}
\label{eq:curl-defn}
(\operatorname{curl} a)_{\square} = 
\sum_{{\bf r}{\bf r}'\in\square}\!\!\!\!\!\!\!\!\!\!\!\circlearrowleft
\; a_{{\bf r}{\bf r'}}\text{.}
\end{equation}

Eq.~(\ref{eq:cubic-gt}) is invariant under gauge transformations written 
in the usual form $a_{{\bf r}{\bf r}'} \rightarrow a_{{\bf r}{\bf r}'} + 
\chi_{{\bf r}'} - \chi_{{\bf r}}$. Therefore $a_{{\bf r}{\bf r}'}$ behaves 
like a vector potential, and since
$[a_{{\bf r}{\bf r}'}, e_{{\bf r}{\bf r}'}] = i$, $e_{{\bf r}{\bf r}'}$ 
plays the role of an electric field. Since the vector potential is a 
$2\pi$-periodic phase this is evidently a \emph{compact} gauge theory. The 
``electric'' charge is given by the lattice divergence of
$e_{{\bf r}{\bf r}'}$:
\begin{equation}
\label{eq:lattice-divergence}
(\operatorname{div} e)_{{\bf r}} =
\sum_{{\bf r}' \leftarrow {\bf r}} e_{{\bf r}{\bf r}'} = \pm 
S^z_{oct}\text{,}
\end{equation}
where the sum is over the sites adjacent to ${\bf r}$. In the ground state 
there is no gauge charge since $S^z_{oct} = 0$, and single gauge charges 
have a large gap of order $J_z$. 

Suppose ${\bf r}$ is a site in the A sublattice. Acting with $S^+_{{\bf 
r},x}$ creates two octahedra with $S^z_{oct} = 1$, at ${\bf r}$ and ${\bf 
r} + {\bf x}$. In the gauge theory language, there is a positive electric 
charge at ${\bf r}$ and a negative one at ${\bf r} + {\bf x}$, where the 
sign of the charge comes from the orientation convention 
Eq.~(\ref{eq:vector-vars-e}). We can now act with the infinite string 
operator:
\begin{equation}
\label{eq:string}
{\cal O}_{string} = \prod_{n = 0}^{\infty} 
S^+_{[{\bf r}+ (n+2){\bf x}+n{\bf y}],y}
S^-_{[{\bf r} + (n+1){\bf x} + n{\bf y}],x} \text{.}
\end{equation}
This hops the gauge charge originally at ${\bf r}+{\bf x}$ off to 
infinity, leaving an isolated $S^z = 1$ octahedron at ${\bf r}$. The two 
$S^z = 1$ octahedra together carry $S^z_{total} = 1$, so the single 
remaining octahedron is evidently a $S^z = 1/2$ \emph{spinon}
\cite{kag,fulde}. The 
spinons are single electric gauge charges and can propagate freely 
in a deconfining phase of the gauge theory; such a state is 
therefore fractionalized. We note that, because spinons cannot hop from
one sublattice to the other, there are in fact two spinon flavors.

Because the theory is compact, magnetic charge is also allowed. Define dual 
lattice sites by serif characters ${\sf r} = {\bf r} + ({\bf x}+{\bf 
y}+{\bf z})/2$; these are located at the centers of the cubic ``boxes'' of 
the direct lattice. The links of the dual lattice are naturally associated 
with the square plaquettes of the direct lattice.
Using this correspondence we define a magnetic field on the dual links
by $\pi b_{{\sf r}{\sf r}'} = (\operatorname{curl} a)_{\square}$.
The sense of the lattice curl 
is taken counterclockwise looking from the point ${\sf r}'$ to ${\sf r}$. The 
lattice divergence $(\operatorname{div} b)_{\sf r}$ gives the magnetic 
charge inside the box at ${\sf r}$. Naively this divergence vanishes since 
each term $a_{{\bf r}{\bf r}'}$ occurs twice, with opposite signs, but 
this is not the case because $b$ is a periodic variable invariant under 
$b \rightarrow b + 2$. In fact we 
have $(\operatorname{div} b)_{\sf r} = 2 n_{\sf r}$ for integer $n_{\sf 
r}$; the magnetic charge is automatically quantized. It is convenient to 
take $b_{{\sf r}{\sf r}'} \in [-1,1)$,
so that $n_{\sf r} = 0,\pm 1,\pm 2$ on each box.  It is also 
possible to have $n_{\sf r} = -3$ in our convention, but this measure 
zero point in the configuration space of magnetic fields should be 
ignored. While electric charges are locally conserved in the 
low-energy manifold, there is no such conservation law for magnetic charge 
and we expect $n_{\sf r} \neq 0$ even in the ground state. While this
means the ground state always has some local 
fluctuations of magnetic charge, it does \emph{not} necessarily contain 
monopoles, which are smoothly-varying defect configurations unaffected by 
a small amount of coarse-graining.

The model Eq.~(\ref{eq:cubic-gt}) looks identical to the 
standard Hamiltonian formulation of compact $U(1)$ lattice gauge theory,
but there is one difference of critical importance in the limit 
of interest $U/K \rightarrow \infty$. Here, the electric field 
takes on half-integer rather than integer values. In the case of integer 
electric fields, the vacuum in the large-$U$ limit is trivial: $e = 0$ 
everywhere with small fluctuations. In our model this limit enforces the 
nontrivial constraint $e = \pm 1/2$. This is another expression of 
the inherent frustration, so we refer to the model as a ``frustrated gauge 
theory.'' This situation is essentially the same as that arising in the large-$N$ limit
of bipartite $SU(N)$ antiferromagnets\cite{read-sachdev1,read-sachdev2}, and in  
the gauge theory description of the square lattice QDM\cite{fradkin-kivelson}. There, one 
considers an integer-valued electric field in a 
background of static charge; of course, our model can also be viewed 
this way by making a shift of the electric field.

To gain insight into the possible phases of ${\cal H}_c$, 
we briefly review the properties of the standard unfrustrated 
gauge theory, which has the same Hamiltonian as our model but 
an integer electric field\cite{kogut}. There are two phases, separated by a 
transition at $(U/K)_c \approx 1$. For $U/K > (U/K)_c$, the ``strong 
coupling'' side of the transition, the model enters a confining phase 
smoothly connected to the trivial $U = \infty$ vacuum $e_{{\bf r}{\bf 
r}'} = 0$. In this phase all excitations are gapped, and static sources of 
the electric field are confined by a linear potential. For $U/K < (U/K)_c$ 
one enters the deconfining Coulomb phase, so named because static gauge 
charges interact via a $1/r$ Coulomb potential and are thus free to propagate. 
At low energies, the effective description of the Coulomb phase is simply 
Gaussian QED, so there is a gapless, linearly 
dispersing photon with two transverse polarizations.

In the Coulomb phase, magnetic monopoles are gapped and interact via a 
$1/r$ magnetic Coulomb potential. The monopoles incorporate the 
compact nature of the magnetic field, which is not important at low 
energies in the deconfined phase. In the confined phase, however, the 
magnetic field fluctuates wildly, its periodicity is important, and 
the monopoles have proliferated and condensed. This distinguishes the two phases: 
in the Coulomb phase the monopole propagator 
decays exponentially, while in the confined phase it goes to a constant. 
This distinction is more robust than the Wilson loop, which fails to 
differentiate between the phases in the presence of matter fields.

Returning to the frustrated gauge theory, we can infer 
that it also has a Coulomb phase at small $U/K$. This should be so because 
in the deconfined phase the discrete nature of the electric field is 
unimportant, so the half-integer nature of $e_{{\bf r}{\bf r}'}$ will not 
play a role. We will be interested in 
whether the Coulomb phase survives in the opposite limit of strong 
coupling, perhaps stabilized by additional terms in the Hamiltonian.

\subsection{Electric-Magnetic Duality}

In recent work on other models with XY ring exchange, it has been useful 
to make a novel ``plaquette duality'' transformation\cite{ebl,triring}. 
In the case of the frustrated gauge theory, plaquette duality is in fact 
identical to the more familiar electric-magnetic duality for $U(1)$ gauge 
theories. We shall make use of the dual cubic lattice defined in the 
preceeding section. We define oriented variables on the dual links 
$\alpha_{{\sf r}{\sf r}'} \in \pi{\mathbb Z}$ and $b_{{\sf r}{\sf r}'} \in 
[-1,1)$, which we take to be canonically conjugate: $[b,\alpha] = i$ on 
the same dual link, zero otherwise. As discussed above, $b_{{\sf r}{\sf 
r}'}$ is the magnetic field, defined by:
\begin{equation}
\label{eq:cubic-magnetic-duality}
\pi b_{{\sf r}{\sf r}'} = (\operatorname{curl} a)_{\square}\text{,}
\end{equation}
where the sense of the curl is again taken counterclockwise looking from
${\sf r}'$ to ${\sf r}$. The conjugate 
variable will play the role of a ``dual vector potential,'' and is related 
to the electric field by:
\begin{equation}
\label{eq:cubic-electric-duality}
\pi (e_{{\bf r}{\bf r}'} - e^0_{{\bf r}{\bf r}'})
= (\operatorname{curl} \alpha)_{\square^*}\text{,}
\end{equation}
where the curl is taken around the \emph{dual} square plaquette encircling the 
direct link joining ${\bf r}$ to ${\bf r}'$, and again the orientation is 
given by the right hand rule. Here $e^0_{{\bf r}{\bf r}'} = \pm 1/2$ is a 
static, divergenceless background field; this is necessary for consistency 
with $e_{{\bf r}{\bf r}'} \in {\mathbb Z} + 1/2$. With these definitions, 
the dual commutation relations are consistent with the original ones.

In dual variables the Hamiltonian is
\begin{equation}
\label{eq:cubic-dual-hamiltonian}
{\cal H}_{cd} = \frac{U}{2 \pi^2} \sum_{\square^*}
\big((\operatorname{curl}\alpha)_{\square^*} + \pi e^0_{{\bf r}{\bf r}'}\big)^2
- K \sum_{\langle{\sf r}{\sf r}'\rangle} \operatorname{cos}(\pi b_{{\sf r}{\sf r}'})
\text{,}
\end{equation}
where the first sum is over dual plaquettes. The constraint of zero 
electric 
charge is now automatically satisfied, since $\operatorname{div} 
e = \operatorname{div}(e^0 + \operatorname{curl} \alpha/\pi) = 0$. 
However, magnetic charge can take on continuous values in the dual 
variables, and one should impose the constraint $\operatorname{div} 
b \in 2{\mathbb Z}$. This constraint commutes with ${\cal H}_{cd}$ 
since it is invariant under the dual gauge transformations $\alpha_{{\sf 
r}{\sf r}'} \rightarrow \alpha_{{\sf r}{\sf r}'} + \lambda_{{\sf r}'} - 
\lambda_{{\sf r}}$, where $\lambda_{{\sf r}} \in \pi {\mathbb Z}$. 
Note that because magnetic charge is not locally conserved, we are not 
allowed to demand $\operatorname{div} b = 0$.

It is useful, and enlightening, to write down the Euclidean action 
obtained by a Trotter expansion in eigenstates of the dual vector 
potential. As usual, one begins with the partition function
\begin{equation}
\label{eq:partition-fn}
Z_{cd} = \operatorname{Tr} \Big(\operatorname{exp}(-\beta {\cal H}_{cd}) 
{\cal P} \Big)\text{.}
\end{equation}
Here ${\cal P} = \prod_{{\sf r}} {\cal P}_{\sf r}$ is a projection 
operator imposing the quantization of magnetic charge:
\begin{eqnarray}
\label{eq:magnetic-quantization-projector}
{\cal P}_{\sf r} &=& \sum_{\alpha_{{\sf r}\tau} \in \pi{\mathbb Z}} 
\delta \big((\operatorname{div} b)_{{\sf r}} - 2 \alpha_{{\sf r}\tau}/\pi 
\big) \\
&=& \frac{1}{2} \sum_{\alpha_{{\sf r}\tau} \in \pi{\mathbb Z}}
\operatorname{exp} \big(i \alpha_{{\sf r}\tau} (\operatorname{div} b)_{\sf 
r} 
\big)\text{.}\nonumber
\end{eqnarray} 

Breaking the exponential in Eq.~(\ref{eq:partition-fn}) into $N_{\tau} = 
\beta / \epsilon$ timeslices, and inserting ${\cal P}$ once in every 
timeslice, one has:
\begin{eqnarray}
\label{eq:trotter}
Z_{cd} &=& \operatorname{Tr} \Big( \operatorname{exp}(-\epsilon {\cal 
H}_{cd}){\cal P} \Big)^{N_\tau} \\
&=& \sum_{\{ \alpha_{{\sf r}\mu} (\tau)\}} \operatorname{exp} 
(-{\cal S}_{cd})\text{,}
\end{eqnarray}
where the index $\mu = \tau,x,y,z$. The imaginary time component of 
$\alpha_{\mu}$ comes from the Poisson-resummed form of the projector 
Eq.~(\ref{eq:magnetic-quantization-projector}). The spatial components 
enter as the eigenvalues of the states $|\{\alpha_{{\sf r}i}(\tau)\}\rangle$ 
used to form the resolution of the identity at each timeslice. 
Following very similar manipulations to those in Appendix A of 
Ref.~(\onlinecite{ebl}), one obtains the dual action:
\begin{eqnarray}
\label{eq:cubic-dual-action}
{\cal S}_{cd} &=& \frac{1}{\pi^2} 
\operatorname{ln}\Big(\frac{2}{\epsilon K}\Big)
\sum_{\tau,\langle{\sf r}{\sf r}'\rangle}
\Big( \Delta_{\tau}\alpha_{{\sf r}{\sf r}'}
+ \alpha_{{\sf r}\tau}
- \alpha_{{\sf r}'\tau} \Big)^2 \nonumber\\
&+& \frac{\epsilon U}{2 \pi^2} \sum_{\tau,\square^*}
\Big( (\operatorname{curl}\alpha)_{\square^*}
+ \pi e^0_{{\bf r}{\bf r}'} \Big)^2 \text{,}
\end{eqnarray}
where $\Delta_{\tau}f \equiv f(\tau + \epsilon) - f(\tau)$.

The action Eq.~(\ref{eq:cubic-dual-action}) is essentially a 
higher-dimensional generalization of the height model partition function 
arrived at by similar manipulations in the context of the square lattice 
QDM\cite{fradkin-kivelson}. Significantly, it differs in having a local rather than a 
global invariance, under spacetime-dependent dual gauge transformations of 
the form $\alpha_{\mu} \rightarrow \alpha_{\mu} + \Delta_{\mu}\lambda$.
In fact, ${\cal S}_{cd}$ has the same structure as non-compact lattice
QED, except for the discrete nature of the fields. This encodes the 
important physics of the magnetic monopoles. In 
Sec.~\ref{sec:effective-theory} below, it will be useful 
to imagine softening the constraint of discreteness on the fields to 
interpolate between the dual partition function and an effective 
description of the Coulomb phase. With the soft constraint (implemented
by the ``corrections'' in Eq.~(\ref{eq:non-ginvt-corrections})), the theory
is identical to a more familiar dual representation of $U(1)$ gauge theory
consisting of a non-compact gauge field minimally coupled to scalar monopoles.

\subsection{Pyrochlore Gauge Theory and Duality}
\label{sec:pyro-mappings}

We now return to the pyrochlore ring model ${\cal H}_p$. In this case the diamond
lattice with sites at the centers of the tetrahedra (discussed in Sec. 
\ref{sec:pyro-model}) plays the role the cubic lattice did for the cubic 
model. Denoting 
\emph{diamond} sites by boldface characters, we soften the hardcore constraint on the 
bosons and go to quantum rotor variables living on the diamond links:
\begin{eqnarray}
\label{eq:pyro-soft}
{\cal H}_p &=& \frac{U}{2} \sum_{\langle{\bf r}{\bf r}'\rangle} (n_{{\bf r}{\bf r}'} - 1/2)^2 \\
 &-& K \sum_{\hexagon}
\operatorname{cos}(\phi_1 - \phi_2 + \phi_3 - \phi_4 + \phi_5 -
\phi_6)\nonumber\text{.}
\end{eqnarray}
Here the second sum is over the hexagonal loops of the diamond lattice (Fig. \ref{fig:diamond}),
and the numbering 
inside the cosine proceeds around the perimeter of the given hexagon. The diamond lattice 
is bipartite, so we define an orientation by declaring that links naturally point 
out of the ``up-pointing'' sites and into the ``down-pointing'' ones (corresponding to up- 
and down-pointing tetrahedra, respectively). We define an oriented electric 
field and vector potential exactly as in 
Eqs.~(\ref{eq:vector-vars-e},\ref{eq:vector-vars-a}). The Hamiltonian then takes the form 
of the diamond lattice frustrated gauge theory:
\begin{equation}
\label{eq:pyro-gt}
{\cal H}_p = \frac{U}{2} \sum_{\langle{\bf r}{\bf r}'\rangle} e^2_{{\bf r}{\bf r}'}
- K \sum_{\hexagon} \operatorname{cos}\Big(
\sum_{{\bf r}{\bf r}'\in\hexagon}\!\!\!\!\!\!\!\!\!\!\!\circlearrowleft
\; a_{{\bf r}{\bf r'}} \Big) \text{.}
\end{equation}
It is evident that the lattice curl now naturally lives on the hexagons of the 
diamond lattice.

Again the electric charge has a simple interpretation in the spin language:
\begin{equation}
\label{eq:pyro-e-charge}
(\operatorname{div} e)_{{\bf r}} = \pm S^z_t \text{.}
\end{equation}
Tetrahedra with $S^z_t = \pm 1$ are now the $S^z = 1/2$ spinons carrying unit gauge 
charge. Single spinons can be created by a string operator similar to 
Eq.~(\ref{eq:string}). For small $U/K$ the model should again enter a deconfining phase 
where the spinons are free to propagate.

We define a dual lattice of plaquette variables by putting a site at the 
center of every pyrochlore hexagon. This is also a pyrochlore lattice, and 
it will be useful to think of its sites as the links of a dual diamond 
lattice with sites labelled by serif characters ${\sf r}$. Each hexagon of 
the dual lattice encircles a link of the direct lattice, and vice versa.
As before, magnetic charge lives on the dual lattice sites. The dual 
variables, again with the commutator $[b,\alpha] = i$ on the same link, 
are defined on the dual links by:
\begin{eqnarray}
\pi (e_{{\bf r}{\bf r}'} - e^0_{{\bf r}{\bf r}'}) &=&
(\operatorname{curl}\alpha)_{\hexagon^*} \\
\pi b_{{\sf r}{\sf r}'} &=& (\operatorname{curl} a)_{\hexagon} \text{,}
\end{eqnarray}
with the sense of the lattice curls determined as in the cubic case. 
Here $(\operatorname{div} b)_{\sf r} = 2n_{\sf r}$, with
$n_{\sf r} = 0,\pm 1$. The Hamiltonian takes the form
\begin{equation}
\label{eq:pyro-dual-hamiltonian}
{\cal H}_{pd} = \frac{U}{2 \pi^2} \sum_{\hexagon^*}
((\operatorname{curl}\alpha)_{\hexagon^*} + \pi e^0_{{\bf r}{\bf r}'})^2
- K \sum_{\langle{\sf r}{\sf r}'\rangle} \operatorname{cos}(\pi b_{{\sf r}{\sf
r}'})\text{.}
\end{equation}
One can derive an action in eigenstates of the dual vector potential as in 
the previous section, with the result:
\begin{equation}
\label{eq:pyro-partition-fn}
Z_{pd} = \sum_{\{\alpha_{{\sf r}{\sf r}'}(\tau)\}}
\sum_{\{\alpha_{{\sf r}\tau}(\tau) \}} \operatorname{exp}(-{\cal 
S}_{pd})\text{,}
\end{equation}
and
\begin{eqnarray}
\label{eq:pyro-dual-action}
{\cal S}_{pd} &=& \frac{1}{\pi^2} 
\operatorname{ln}\Big(\frac{2}{\epsilon K}\Big)
\sum_{\tau,\langle{\sf r}{\sf r}'\rangle}
\Big( \Delta_{\tau}\alpha_{{\sf r}{\sf r}'}
+ \alpha_{{\sf r}\tau}
- \alpha_{{\sf r}'\tau} \Big)^2 \nonumber\\
&+& \frac{\epsilon U}{2 \pi^2} \sum_{\tau,\hexagon^*}
\Big( (\operatorname{curl}\alpha)_{\hexagon^*}
+ \pi e^0_{{\bf r}{\bf r}'} \Big)^2 \text{.}
\end{eqnarray}
The constraint of magnetic charge quantization $(\operatorname{div} 
b)_{{\sf r}} \in 2{\mathbb Z}$ enters as before, giving rise to the 
temporal dual vector potential fields in $Z_{pd}$.

\section{Effective Theory}
\label{sec:effective-theory}

\subsection{Coulomb Phase Effective Action}
It is well-known that the low-energy description of the Coulomb phase of 
compact $U(1)$ gauge theory is non-compact QED 
with no matter fields. In our case it will be convenient to work 
in dual variables to formulate the effective theory, which more naturally 
allows 
the inclusion of magnetic charge fluctuations. Depending on the purpose at 
hand, different formulations of the effective action will be useful. To 
fix notation, we show them all here. In each case the partition function 
is of the form:
\begin{equation}
\label{eq:general-dual-pf}
{\cal Z}^0 = \prod_{\tau} \prod_{{\sf r}, \mu}
\int [d\tilde{\alpha}_{{\sf r},\mu}(\tau)]
\operatorname{exp}(-{\cal S}^0)\text{,}
\end{equation}
where $\tilde{\alpha}$ is a real-valued field.
It will often be convenient 
to keep the full spatial lattice structure, in which case we can define 
$\tilde{\alpha}$ in terms of the microscopic variables by the
\emph{temporal} coarse-graining:
\begin{equation}
\label{eq:temporal-coarse-graining}
\tilde{\alpha}_{{\sf r}\mu} = [\alpha_{{\sf r}\mu}]_f -
\frac{\alpha^0_{{\sf r}\mu}}{2}\text{.}
\end{equation}
The brackets $[\,]_f$ denote an average over high-frequency modes
in imaginary time, and we have subtracted the time-independent, 
non-fluctuating background
$\alpha^0_{{\sf r}{\sf r}'} \in \pi {\mathbb Z}$ defined by:
\begin{equation}
\label{eq:alpha0-defn}
\frac{1}{\pi} (\operatorname{curl} \alpha^0)_{\square^*} =
- 2 e^0_{{\bf r}{\bf r}'}\text{,}
\end{equation}
with $\tilde{\alpha}^0_{{\sf r}\tau} = 0$.
If (as we always do in practice) we restrict attention to spatially periodic
$e^0_{{\bf r}{\bf r}'}$ with zero average electric flux in every
direction, $\alpha^0$ can also be taken periodic. This subtraction
simplifies the relation between the electric field and dual vector
potential, since
\begin{equation}  
\label{eq:simpler-duality}
\pi e_{{\bf r}{\bf r}'} = \big( \operatorname{curl} (\alpha -
\alpha^0/2)\big)_{\square^*}\text{.}
\end{equation}

On a spacetime lattice, the action looks almost identical to ${\cal 
S}_{cd}$ in Eq.~(\ref{eq:cubic-dual-action}):
\begin{eqnarray}
\label{eq:discrete-cubic-effective-theory}
{\cal S}^0_{lat} &=& \frac{g_{\tau}}{2} \sum_{\tau}
\sum_{\langle{\sf r}{\sf r}'\rangle} 
(\Delta_{\tau} \tilde{\alpha}_{{\sf r}{\sf r}'} +
\tilde{\alpha}_{{\sf r}\tau} - \tilde{\alpha}_{{\sf r}'\tau})^2 \\
&+& \frac{g_s}{2} \sum_{\tau}\sum_{\square^*} (\operatorname{curl} 
\tilde{\alpha})^2 
\nonumber\text{.}
\end{eqnarray}
The only difference from the microscopic model is that the fields are now 
continuous. We will also have occasion to retain the 
spatial lattice structure but take the time-continuum limit, in which case 
we write
\begin{eqnarray}
\label{eq:time-cont-cubic-effective-theory}
{\cal S}^0_{tc} &=& \frac{1}{2{\cal K}} \int d\tau
\sum_{\langle{\sf r}{\sf r}'\rangle}
(\partial_{\tau} \tilde{\alpha}_{{\sf r}{\sf r}'} +
\tilde{\alpha}_{{\sf r}\tau} - \tilde{\alpha}_{{\sf r}'\tau})^2 \\
&+& \frac{{\cal U}}{2} \int d\tau \sum_{\square^*} 
(\operatorname{curl}\tilde{\alpha})^2 \nonumber\text{.}
\end{eqnarray}
The parameters here are related to those in 
Eq.~(\ref{eq:discrete-cubic-effective-theory}) by
${\cal K} = (\epsilon g_{\tau})^{-1}$ and ${\cal U} = g_s/\epsilon$, and 
$\tilde{\alpha}_{{\sf r}\tau}$ now has units of inverse time. This action corresponds
to
the effective dual Hamiltonian (in the sector with no magnetic charge):
\begin{equation}
\label{eq:cubic-effective-hamiltonian}
{\cal H}^0 = \frac{{\cal K}}{2}
\sum_{\langle{\sf r}{\sf r}'\rangle} \tilde{b}^2_{{\sf r}{\sf r}'} 
+ \frac{{\cal U}}{2} \sum_{\square^*} 
(\operatorname{curl}\tilde{\alpha})^2 \text{,}
\end{equation}
which can be obtained simply by expanding the cosine in 
Eq.~(\ref{eq:cubic-dual-hamiltonian}). The lattice effective actions for 
the pyrochlore model look identical to 
Eqs.~(\ref{eq:discrete-cubic-effective-theory},
\ref{eq:time-cont-cubic-effective-theory}), where the only change 
necessary is the replacement of the sums over dual square plaquettes with 
sums over dual hexagons.

Finally, in order to take the spatial continuum limit, we introduce a 
continuum 4-vector field $\Upsilon_{\mu}({\bf r},\tau)$, formally 
defined by the replacements 
\begin{eqnarray}
\label{eq:biga-defn}
\tilde{\alpha}_{{\sf r}i} &\to& l ({\bf e}_i \cdot {\bf \Upsilon}) \\
\tilde{\alpha}_{{\sf r}\tau} &\to& \Upsilon_{\tau} \nonumber \text{.}
\end{eqnarray}
Here ${\bf e}_i$ are the vectors connecting the site ${\sf r}$ to its 
nearest neighbors, in either the cubic or diamond lattice, and $l$ is a microscopic
length on the order of the lattice spacing.
Na\"{\i}vely 
it seems we have thrown away too much information in the pyrochlore 
case, since there are four sites per unit cell, but only three spatial 
components of the continuum vector field. However, there are also 
\emph{two} gauge degrees of freedom per unit cell of the diamond lattice, 
corresponding to changing $(\operatorname{div} \alpha)_{\sf r}$ on the two 
different sublattices. This leaves us with two transverse degrees of 
freedom, the same number as in the cubic model and the continuum 
theory. The full spacetime continuum theory is
\begin{eqnarray}
\label{eq:dual-continuum-action}
{\cal S}^0_{stc} &=& \frac{1}{2{\cal K}_c} \sum_i \int d\tau d^3{\bf r}
(\partial_{\tau}\Upsilon_i - \partial_i  
\Upsilon_{\tau})^2 \\
&+& \frac{{\cal U}_c}{2} \sum_{i < j} \int d\tau d^3{\bf r}
(\partial_i \Upsilon_j - \partial_j \Upsilon_i)^2 
\nonumber \text{,}
\end{eqnarray}
where ${\cal K}_c = {\cal K} l$ and ${\cal U}_c = {\cal U} l$
($l$ is the microscopic length in Eq.~\ref{eq:biga-defn}). The spatial/temporal components of 
$\Upsilon_{\mu}$ have units of inverse length/time. The form is identical to the 
familiar (dual) Maxwell action for electromagnetism with photon velocity 
$v_p = \sqrt{{\cal U}_c {\cal K}_c}$. This action can be used to 
obtain long-wavelength properties of the Coulomb phase of either 
microscopic model, but at the end of any calculation the allowed spatial 
components of all vector fields are determined by the lattice structure.

To calculate using any of these effective actions, it is convenient
to implement a gauge-fixing procedure. The well-known manipulations
of Faddeev and Popov\cite{f-p} tell us that we may add any function of the vector potential
4-divergence to the action. The standard choice, in our nonstandard notation, is:
\begin{equation}
\label{eq:fp-action}
{\cal S}_{FP} = \frac{1}{2\xi\,{\cal U}_c{\cal K}_c^2}
\int d\tau d^3{\bf r} (\partial_{\tau} \Upsilon_{\tau} +
v_p^2 {\bf \nabla}\cdot{\bf \Upsilon})^2 \text{.}
\end{equation}
If we choose $\xi = 1$, as usual the off-diagonal terms in the action are cancelled and
we obtain the simple photon propagator:
\begin{equation}
\label{eq:photon-propagator}
\langle \Upsilon_{\mu}({\bf k},\omega_n) \Upsilon_{\nu}({\bf k}',\omega_n') \rangle = 
\frac{(2\pi)^4 \delta(\omega_n + \omega_n')\delta({\bf k} + {\bf k}') 
{\cal K}_c g_{\mu \nu}}{\omega_n^2 + v_p^2 k^2}\text{,}
\end{equation}
where the Euclidean metric is defined by $g_{\tau\tau} = v_p^2$,
$g_{\tau i} = g_{i \tau} = 0$ and $g_{ij} = \delta_{ij}$. While these expressions hold in the
continuum, it is amusing to note that
the obvious lattice regularization of Eq.~(\ref{eq:fp-action}) also cancels all off-diagonal
terms on the cubic lattice. Working in the time-continuum action, we find
the lattice propagator:
\begin{equation}
\label{eq:lattice-photon-propagator}
\langle \tilde{\alpha}_{\mu}({\bf k},\omega_n) \tilde{\alpha}_{\nu}({\bf k}',\omega_n')
\rangle =
\frac{(2\pi)^4 \delta(\omega_n + \omega_n')\delta({\bf k} + {\bf k}') 
{\cal K} g_{\mu \nu}}{\omega_n^2 + {\cal U}{\cal K}g({\bf k})}\text{,}
\end{equation}
where $g({\bf k}) = 2\sum_i (1 - \operatorname{cos}k_i)$. On the diamond lattice, even more
complicated Faddeev-Popov terms seem only to cancel some of the off-diagonal terms.

Finally, it is important to note that this procedure is only legitimate for calculating
expectation values of operators invariant under \emph{continuous} dual gauge transformations.
Terms with only discrete dual gauge invariance (see Sec.~\ref{sec:corrections}) must be
handled more carefully; this issue arises in calculating the monopole propagator in
Section \ref{sec:monopole-prop}.

\subsection{RK Point Effective Action}

Given the microscopic gauge structure of our models, it is reasonable to conjecture that
the low-energy effective degrees of freedom are simply noncompact $U(1)$ gauge fields. 
The effective action should take the form of an expansion in the lowest-order terms 
involving the dual vector potential consistent with the symmetries. Generically this would take
the form of Eq.~(\ref{eq:dual-continuum-action}); however, as our analysis of the microscopic models
takes advantage of the special properties of the Rokhsar-Kivelson point to conclude that the ``stiffness''
or ${\cal U}_c$ term
for the electric field vanishes there, we must include a higher term.
This leads us to propose that the RK point of both models is described by the effective action
\begin{eqnarray}
\label{eq:rk-effective-action}
{\cal S}^0_{RK} &=& \frac{1}{2{\cal K}_c} \sum_i \int d\tau d^3{\bf r} 
(\partial_{\tau}\Upsilon_i - \partial_i  
\Upsilon_{\tau})^2 \\
&+& \frac{{\cal W}_c}{2} \int d\tau d^3{\bf r}
\big( {\bf \nabla} \times ({\bf \nabla} \times {\bf \Upsilon})\big)^2 
\nonumber \text{.}
\end{eqnarray}
Here ${\cal W}_c$ is the coefficient of the new term, which (when ${\cal U}_c = 0$) is marginal in the
renormalization group sense and must be included.
While the theory remains
gauge invariant, the photon dispersion now vanishes \emph{quadratically}:
$\omega \sim {\bf k}^2$.
The ${\cal U}_c$ term is relevant, as can be seen by the usual power-counting procedure, so
this action cannot describe a stable phase. A small positive ${\cal U}_c$ will drive the
system into the Coulomb phase, while a small negative ${\cal U}_c$ will result in an
electric field ``crystal;'' in the dimer language this is a state with long-range dimer order
and no flippable plaquettes.
As we argue in Section \ref{sec:rkpoint}, precisely 
this picture describes the physics of the microscopic models near the RK point, with the
relation ${\cal U}_c \propto (1 - V/J_{ring})$.

A very similar story is known to apply to the $d=2$ square lattice QDM
\cite{henley-unpub}. In that case, electric-magnetic duality leads to a height
model partition function, which describes the fluctuations of a discrete field living
on the square lattice plaquettes in 2+1 dimensions. Just as above, softening the constraint
of discreteness naturally leads to the Gaussian effective action:
\begin{equation}
\label{eq:height-effective}
{\cal S}_{height} = \int d\tau d^2{\bf r} \Big( (\partial_{\tau} h)^2
+ \kappa_1 ({\bf \nabla} h)^2 + \kappa_2 (\nabla^2 h)^2 \Big)\text{.}
\end{equation}
When $\kappa_1 = 0$ we have a description of the RK point, which has power-law dimer-dimer
correlations. For small negative $\kappa_1$ the system goes into the staggered valence
bond crystal state. These have natural analogs in the three-dimensional case. 
However, a small
positive $\kappa_1$ leads to a confining state with broken translation
symmetry\cite{square-diag, early-square-diag} via an
instability of the Gaussian theory.  This occurs because
Eq.~(\ref{eq:height-effective}) with $\kappa_1 > 0$ would describe the unstable
Coulomb phase of pure 2+1-dimensional $U(1)$ gauge theory. In three dimensions, however,
the analogous phase is stable and should exist adjacent to the RK point.

To calculate with Eq.~(\ref{eq:rk-effective-action}) we again carry out a gauge-fixing
procedure. We add the Faddeev-Popov term:
\begin{equation}
\label{eq:rk-fpterm}
{\cal S}_{FP-RK} = \frac{1}{2\xi{\cal K}^3_c} \int d\tau d^3{\bf r}
(\partial_{\tau}\Upsilon_{\tau} + {\cal K}_c^2 {\bf \nabla}\cdot{\bf \Upsilon})^2 \text{.}
\end{equation}
Although this does not cancel all off-diagonal terms, it leads to the relatively
simple photon propagator:
\begin{eqnarray}
\label{eq:rk-photon-prop}
&\langle& \!\!\!\!\Upsilon_{\mu}({\bf k},\omega_n) \Upsilon_{\nu}({\bf k}',\omega_n') \rangle \\ 
&=& (2\pi)^4 \delta(\omega_n + \omega_n') \delta({\bf k} + {\bf k}') {\cal K}_c
{\cal M}_{\mu \nu}({\bf k},\omega_n) \text{,}\nonumber
\end{eqnarray}
where
\begin{eqnarray}
\label{eq:rk-photon-prop-matrix}
{\cal M}_{\tau\tau} &=& {\cal K}_c^2 /(\omega_n^2 + {\cal K}_c^2 k^2) \\
{\cal M}_{\tau i} &=& {\cal M}_{i \tau} = 0 \nonumber\\
{\cal M}_{i i} &=& \frac{1}{f(\omega_n,{\bf k})} 
\big[ m^2 \omega_n^2 + k_i^2 k^2 + m^2 {\cal K}_c^2 (k^2 - k_i^2) \big] \nonumber\\
{\cal M}_{i j} &=& \frac{1}{f(\omega_n,{\bf k})}
\big[ k_i k_j (k^2 - {\cal K}_c^2 m^2) \big] \quad \text{(}i \neq j\text{)}\nonumber\text{,}
\end{eqnarray}
and
\begin{equation}
\label{eq:rk-photon-prop-function}
f(\omega_n,{\bf k}) = (\omega_n^2 + {\cal K}_c^2 k^2)(m^2 \omega_n^2 + k^4) \text{,}
\end{equation}
with $m = ({\cal K}_c {\cal W}_c)^{-1/2}$.

\subsection{Corrections}
\label{sec:corrections}

While all gauge-invariant
corrections to ${\cal S}^0_{stc}$ are irrelevant in the renormalization group
sense, in order to understand the detailed realization of the $U(1)$ spin 
liquid (and the nearby RK point) in any microscopic model it will generally be important
to consider corrections to
the effective action. Furthermore, we need to specify 
relations between the microscopic and effective degrees of freedom. 
Na\"{\i}vely we could write $\alpha_{\mu} \sim \tilde{\alpha}_{\mu} 
+ \alpha^0_{\mu}/2$, but any corrections allowed by symmetry will 
generically be present.

For ease of presentation, we focus on the cubic model; the lattice
structure is unimportant for these results. The symmetries of the
microscopic model are listed in detail in Appendix \ref{app:symmetries}. It is important
to note that, because the background $\alpha^0$ is not invariant under all
lattice symmetries, the dual vector potential transforms with additional shifts under
these operations. Furthermore, the dual effective degrees of freedom $\tilde{\alpha}$ and
$\tilde{b}$ transform exactly as their microscopic counterparts.

Rather than attempt a painstaking enumeration of all allowed corrections to ${\cal S}^0$,
we discuss some representative examples. The corrections naturally fall into two classes;
those invariant under \emph{continuous} dual gauge transformations, and those not. In
Hamiltonian language on the lattice, some typical terms in the first class are:
\begin{eqnarray}
\label{eq:ginvt-corrections}
{\cal H}^1 &=& \frac{{\cal W}}{2} \sum_{\square}(\operatorname{curl} \tilde{e})^2
+ {\cal W}' \sum_{\square^*} (\operatorname{curl} \tilde{b})^2 \\
&+& {\cal U}' \sum_{\langle{\bf r}{\bf r}'\rangle} \tilde{e}^4_{{\bf r}{\bf r}'}
+ {\cal K}' \sum_{\langle{\sf r}{\sf r}'\rangle} \tilde{b}^4_{{\sf r}{\sf r}'}
+ \cdots\nonumber
\end{eqnarray}
Terms involving discrete line integrals of both vector potentials around closed
loops larger than single plaquettes are also allowed, as are terms containing the
divergence of the fields. While these terms are irrelevant,
they can have quantitative effects, presumably accessible in perturbation theory. All such
corrections
are also irrelevant at the RK point, except the ${\cal W}$ term, which contributes to the
${\cal W}_c$ term in Eq.~(\ref{eq:rk-effective-action}).

More interesting are those terms lacking continuous dual gauge invariance. 
Here we only
consider the single site terms; in the spacetime lattice action these take the form:
\begin{eqnarray}
\label{eq:non-ginvt-corrections}
{\cal S}^2 &=&
-\sum_{\tau,\langle{\sf r}{\sf r}'\rangle} \sum_{q = 1}^{\infty}
v_{2q} \operatorname{cos}^q (\alpha^0_{{\sf r}{\sf r}'})
\operatorname{cos}(2 q \tilde{\alpha}_{{\sf r}{\sf r}'}) \\
&-& \sum_{\tau,{\sf r}} \sum_{q=1}^{\infty} v^{\tau}_{2q}
\operatorname{cos}(2 q \tilde{\alpha}_{{\sf r}\tau}) \text{.}\nonumber
\end{eqnarray}
In the first term the $\alpha^0$-dependence is necessary to compensate the shifts
in $\tilde{\alpha}$ under lattice symmetries. With appropriate choices of coefficients,
when these corrections become large they have the effect of pinning the dual vector
potential to take on discrete values. This allows us to interpolate
explicitly between the effective theory ${\cal S}^0_{lat}$
and the microscopic partiton function. Physically,
the spatial $v_{2q}$ part of ${\cal S}^2$ is a \emph{magnetic charge hopping} term. Its temporal
companion is related to the discreteness of magnetic charge. Therefore, these corrections
introduce magnetic charge fluctuations into the effective theory. As should be expected
when magnetic monopoles are gapped, when these fluctuations are small there is no
associated instability. Formally, this should be the case because the correlation functions
of $\operatorname{cos}(2 q \tilde{\alpha}_{\mu})$ are local in space and time. For example,
\begin{equation}
\label{eq:local-correlator}
\Big\langle \operatorname{cos}\big(2 \tilde{\alpha}_{{\sf r}x}(\tau)\big)
\operatorname{cos}\big(2 \tilde{\alpha}_{{\sf r}'x}(\tau')\big)\Big\rangle_0 
\propto \delta_{{\sf r}{\sf r}'} \delta_{\tau \tau'}\text{,}
\end{equation}
because continuous dual gauge invariance in ${\cal S}^0$ makes each cosine into an
independent random variable. 

Finally, we consider corrections to the scaling equalities 
between microscopic and effective
degrees of freedom. Some representative contributions in the case of the
dual vector potential are, in operator language:
\begin{eqnarray}
\label{eq:dvp-corrections}
\alpha_{{\sf r}{\sf r}'} &\sim&
(\tilde{\alpha}_{{\sf r}{\sf r}'} + \alpha^0_{{\sf r}{\sf r}'}/2) \\
&+& \sum_{q = 1}^{\infty} c_{2q} \operatorname{cos}^q (\alpha^0_{{\sf r}{\sf r}'})
\operatorname{sin}(2 q\tilde{\alpha}_{{\sf r}{\sf r}'}) \nonumber\\
&+& \sum_{n,m = 0}^{\infty} c^1_{n,m} (\tilde{b}_{{\sf r}{\sf r}'})^{2n}
(\operatorname{curl} \tilde{e})_{\square}^{2m+1} + \cdots\nonumber
\end{eqnarray}
This should really be interpreted as an instruction for what effective
theory operators to use in the calculation of gauge invariant expectation
values. These corrections are not important, and we will neglect them:
the second term clearly only leads to local corrections,
while the third term and others like it lead only to subdominant power laws,
which will always be present on the lattice. The conclusion is the same for
corrections to the magnetic field, but for completeness we show some of the
representative terms:
\begin{eqnarray}
\label{eq:magnetic-corrections}
e^{i b_{{\sf r}{\sf r}'}} \sim \operatorname{exp}\Big(&i& \tilde{b}_{{\sf r}{\sf r}'} \\
+ &i& \sum_{n,m=0}^{\infty} d^1_{n,m} (\operatorname{curl}\tilde{e})^{2n}_{\square}
(\tilde{b}_{{\sf r}{\sf r}'})^{2m+1} + \cdots\Big) \nonumber\\
\cdot\;\Bigg\{&1& +
\sum_{q=1}^{\infty} d_{2q} \operatorname{cos}^q (\alpha^0_{{\sf r}{\sf r}'})
\operatorname{cos}(2q\tilde{\alpha}_{{\sf r}{\sf r}'}) +
\cdots \Bigg\} \nonumber
\end{eqnarray}

\section{Properties of the $U(1)$ Spin Liquid}
\label{sec:properties}

Using the effective theory developed above, we now treat some of
the more striking properties of the $U(1)$ spin liquid. Since the effective theories
for both the generic Coulomb phase and the adjacent RK point are quadratic in the fields, 
all of the calculations can be done by simple Gaussian integrals.

\subsection{Excitations and Emergent Long-Range Interactions}

As already mentioned, the $U(1)$ spin liquid supports a gapped $S^z = 1/2$ spinon
carrying electric gauge charge, a gapped topological point 
defect that plays the role of a magnetic monopole, and a gapless, linearly
dispersing ``photon'' with velocity $v_p$. The spinon was discussed in Section
\ref{sec:frust-gt}, and in the microscopic models takes the form of a tetrahedron
(or octahedron) carrying nonzero $S^z$. The gap to the spinons is very large, of order $J_z$.

The monopole is a classical configuration of the magnetic field emanating from a point
with nonzero magnetic charge at ${\sf r}_0$.
For a compact gauge field on, say, the cubic lattice,
the classical configuration ${\cal A}$ of the vector potential is given by minimizing
\begin{equation}
\label{eq:monopole-classical-energy}
E[{\cal A}] = -\sum_{\langle{\sf r}{\sf r}'\rangle} 
\operatorname{cos}(\pi{\cal B}_{{\sf r}{\sf r}'})
\end{equation}
in the presence of a specified distribution of magnetic
charge, which enters via the constraint
$(\operatorname{div} {\cal B})_{{\sf r}} = 2 \delta_{{\sf r}{\sf r}_0}$.
In these expressions $\pi{\cal B}_{{\sf r}{\sf r}'} = (\operatorname{curl} {\cal A})_{\square}$.
Far from the center of the monopole the magnetic field will be small, the energy is
approximately a sum over ${\cal B}^2_{{\sf r}{\sf r}'}$, and the minimum is obtained
by solving the (discrete) Laplace equation of classical electrostatics.
The compact nature of the theory only modifies the field near the core.

In the effective theory, we can create a monopole with the operator
\begin{equation}
\label{eq:effective-monopole-creation}
\tilde{m}^{\dagger}_{{\sf r}_0} = \operatorname{exp}(i \sum_{\langle{\sf r}{\sf r}'\rangle}
{\cal B}_{{\sf r}{\sf r}'} \tilde{\alpha}_{{\sf r}{\sf r}'})\text{.}
\end{equation}
The exact choice of ${\cal B}$ in this operator is somewhat arbitrary; the requirement
for monopole creation is that the surface integral of the magnetic field at large distances
indicates the presence of one magnetic charge. This arbitrariness does not matter when
calculating long-time monopole correlators, which will be dominated
by the contribution from the lowest-energy monopole eigenstate (it is not hard to see that,
in the effective theory, our operator has very good overlap with this state). As long
as the magnetic field is chosen to spread the flux uniformly from the monopole center,
long-distance correlations should also be unaffected.
Because magnetic charge is created at ${\sf r}_0$, this operator is \emph{not} invariant
under continuous dual gauge transformations. The monopole gap is of order ${\cal K}$, and
in the microscopic models is likely to be of order $J_{ring}$ as long as ${\cal K}$ does not
renormalize too much from its bare value. Therefore the monopoles have a much smaller gap than
the spinons.

Using the correspondence between microscopic and effective variables, we can write down
a creation operator that should have at least some overlap with the true microscopic monopole
eigenstate:
\begin{equation}
\label{eq:micro-monopole-creation}
m^{\dagger}_{{\sf r}_0} = \operatorname{exp}(i \sum_{\langle{\bf r}{\bf r}'\rangle}
{\cal A}_{{\bf r}{\bf r}'} e_{{\bf r}{\bf r}'})\text{.}
\end{equation}
This is written in terms of the direct variables to connect with the spin
language, where the monopoles are defect configurations in the XY component
of the spin. Because of the half-integer electric field,
the sign of $m^{\dagger}$ is changed by a $2\pi$-shift of any one of the
${\cal A}_{{\bf r}{\bf r}'}$. However, this overall sign does not fluctuate since
the electric fields change only by integer steps.

Both the monopoles and spinons feel an emergent $1/r$ interaction, even though
the microscopic Hamiltonians contain only local operators. Consider the field due to
a configuration of a few static electric gauge charges (the same discussion could be
repeated for magnetic charge).
This is given by solving Poisson's equation $\nabla \cdot {\bf E}
= \sum_i q_i\delta({\bf r}-{\bf r}_i)$ to obtain Coulomb's Law. By
gauge invariance the longitudinal part of ${\bf E}$ does not fluctuate, so
the field is simply given by its classical value plus \emph{transverse} fluctuations. This zero
point energy does not contribute to the energy difference between the charged
state and the charge-free ground state, and one recovers the familiar $1/r$ Coulomb
potential. In the RK point effective theory, this interaction is absent for
the spinons, but still present for monopoles,
because there is no energy cost for longitudinal electric fields.

A word about the statistics of the charged excitations is in order. Certainly,
both monopoles and electric charges as discussed above are bosonic. However,
a simple argument shows that a \emph{dyon}, a bound state of electric and magnetic
charge, has Fermi statistics\cite{dyons}. If such bound states exist and have lower energy
than pure electric charges, then one could say that there are fermionic spinons. 
While this is a nonuniversal question with no bearing on the low-energy physics, it can be
relevant in understanding the possible transitions out of the $U(1)$ spin liquid. 
The emergence
of Fermi statistics in local bosonic models is also of great conceptual interest\cite{malevin},
although in the present case it is not clear what, if any,
local microscopic interaction will bind electric and magnetic charge.

Finally we turn to the photon. In terms of the
spins, it is a linearly dispersing
$S^z = 0$ collective mode oscillating between the Ising and XY parts of the spin
vector. As for \emph{phonons}, these gapless excitations
make the contribution to the
low-temperature specific heat $C_{photon}(T) \propto (T/v_p)^3$. Since in real materials
it should be possible to quantitatively understand the phonon contribution to $C(T)$ by
measurement of elastic moduli\cite{broholm}, this 
signature of the $U(1)$ spin liquid
should be easily accessible to experiments. Other potential, though likely more delicate,
probes of the photon are low-temperature thermal conductivity and Raman scattering. 
Finally, the photon manifests itself in the power-law correlations
discussed in Section \ref{sec:power-laws}.

\subsection{$U(1)$ Topological Order}
\label{sec:toporder}

In the theory of $Z_2$-fractionalized phases, the notion of topological ground state
degeneracy\cite{wen-niu} has been very useful both as a conceptual tool, and as a property that can
be directly probed by experiments\cite{sf-senthil-expt,sf-topord,visonexp1,visonexp2}.
The degeneracy is associated with
the topological sectors of a deconfining $Z_2$ gauge field, so it is natural to ask about the
generalization of these ideas to $U(1)$-fractionalized states.

We will work with the microscopic 
Hamiltonian of cubic frustrated gauge theory (Eq.~\ref{eq:cubic-gt})
on a 3-torus, a cube with $L$ lattice sites on a side and periodic boundary conditions:
\begin{eqnarray}
\label{eq:top-bc}
e_{[{\bf r}+L{\bf e}_i],j} &=& e_{{\bf r}j} \\
\operatorname{exp}(i a_{[{\bf r}+L{\bf e}_i],j}) &=& \operatorname{exp}(i a_{{\bf r}j})
\nonumber \text{.}
\end{eqnarray}
We define an operator to measure the electric flux
through a plane perpendicular to each of the three independent cycles of the torus. For
example,
\begin{equation}
\label{eq:eflux-defn}
\Phi^E_x = \sum_{n_y,n_z=0}^{L-1} e_{[n_y{\bf y}+n_z{\bf z}],x} \text{,}
\end{equation}
with similar definitions for $\Phi^E_{y,z}$. These fluxes are 
constants of the motion, and are in fact conserved by \emph{any} local, gauge invariant
dynamics. Furthermore, because there is no electric charge in the ground
state, by Gauss' Law we can make arbitrary incremental deformations of the specified plane 
without changing the flux. These fluxes define the electric topological sectors
of the gauge theory\cite{sectors}. 

Now imagine the model is in the Coulomb phase,
where in the ground state $\Phi^E_i = 0$. We imagine
threading one quantum $\Phi^E_0$ of electric flux 
through the system in, say, the $x$-direction. Because the energy is proportional to
$\int_{\bf r} {\bf E}^2$, the most favorable situation
is for the flux to spread itself uniformly through the system, so that $E_x = \Phi^E_0/L^2$. This
electric field is purely longitudinal and does not fluctuate, so there is a total energy cost
${\cal E} \sim {\cal U}_c(\Phi^E_0)^2/L$. Therefore, in the thermodynamic limit, all states with
a finite number of electric flux quanta threading each direction of the 3-torus become degenerate
with the ground state. Just as in the case of $Z_2$ topological order, these states are locally
identical to the ground state (local correlation functions will be unaffected), but globally
distinct.

What about \emph{magnetic} topological sectors of the gauge field? Because 
$[b_{{\sf r}{\sf r}'},\Phi^E_i] = 0$, these can clearly be specified simultaneously
with the electric sectors. We define magnetic fluxes $\Phi^B_i$ just as above:
\begin{equation}
\label{eq:bflux-defn}
\Phi^B_x = \sum_{n_y,n_z=0}^{L-1} b_{[n_y{\sf y}+n_z{\sf z}],x} \text{,}
\end{equation}
and similarly for $\Phi^B_{y,z}$.
This expression is a (lattice) surface integral over a plane bisecting the system,
which has the topology of a 2-torus due to the periodic boundary conditions.
Because this is a closed surface and $b$ is the curl of a compact vector potential,
the magnetic flux is quantized: $\Phi^B_i = 2n^B_i$.

\begin{figure}
\begin{center}
\centerline{\includegraphics[width=3.5in]{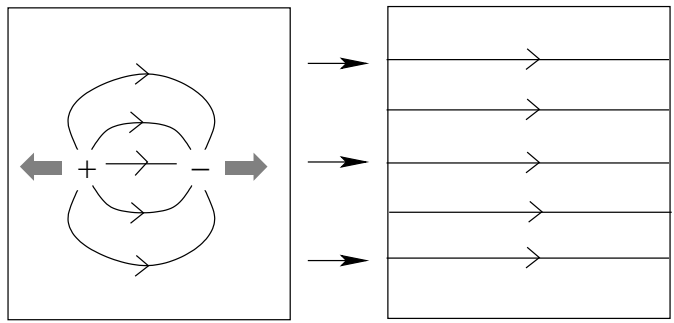}}
\caption{Illustration of threading a magnetic flux quantum through one direction
of a periodic system by creating a monopole-antimonopole pair (left), and
moving it apart until the charges return to the same position and 
annihilate. One is left with no monpoles, and magnetic flux threading the system (right).}
\label{fig:flux-thread}
\end{center}
\end{figure}

However, there is an important difference from
the case of electric flux: because magnetic charge fluctuates in the ground state,
$\Phi^B_i$ is \emph{not} a constant of the motion in the microscopic theory. Within the
Coulomb phase, though, the magnetic flux defined
in the effective theory does commute with ${\cal H}^0$, the effective Hamiltonian written
in dual variables. Magnetic charge fluctuations can mix states with different values of
the flux by creating a monopole-antimonopole pair and separating the particles along one lattice
direction, only to have them annihilate when they complete their periodic journey around
the system (Fig.~\ref{fig:flux-thread}).
Because the monopoles are gapped, the rate for this process is
suppressed exponentially in the system size, and it does not occur for a large enough sample.
As with the electric sectors, the magnetic flux sectors become degenerate ground states
as $L \to \infty$ with energies scaling to zero as $1/L$. In the microscopic theory, we
expect that $\langle \Phi^B_i \rangle$ will label the topological sectors within the Coulomb
phase; in confining phases magnetic charge fluctuates wildly and $\Phi^B_i$ is no longer
a good quantum number, even approximately.

To summarize, the $U(1)$ spin liquid on a 3-torus
has a topological degeneracy characterized by six \emph{integers}, with energies that vanish as
\begin{eqnarray}
\label{eq:top-sector-energy}
{\cal E} &\sim& {\cal U}_c \frac{(\Phi^E_0)^2}{L} \Big[(n^E_x)^2 + (n^E_y)^2 + (n^E_z)^2 \Big]\\
&+& {\cal K}_c \frac{(\Phi^B_0)^2}{L} \Big[(n^B_x)^2 + (n^B_y)^2 + (n^B_z)^2 \Big]\nonumber
\text{.}
\end{eqnarray}
In the RK point effective action ${\cal U}_c$ vanishes, so the electric topological sectors
have zero energy, with only corrections from irrelevant terms possible.

\subsection{Spin-Spin and Plaquette-Plaquette Correlators}
\label{sec:power-laws}

We first consider the two-point $S^z$ correlation function both in the Coulomb
phase and at the RK point; the transverse
part of the spin-spin correlator is not gauge-invariant and vanishes.
The mappings of
Section \ref{sec:frust-gt} tell us this is given by the electric field
correlator. In the cubic case we
use the three-site unit cell containing $\{{\bf r}x,{\bf r}y,{\bf r}z\}$.
Because of the orientation convention for the electric field, we
have $S^z_{{\bf r}i} = \operatorname{exp}(i {\bf K}_0 \cdot {\bf r}) e_{{\bf r}i}$, where
${\bf K}_0 = (\pi,\pi,\pi)$ and we take ${\bf r} = 0$ in the $A$ sublattice. Therefore
the spin correlation function will be shifted from the electric field one by ${\bf K}_0$
in the Brillouin zone. For the pyrochlore we use the unit cell of
Section \ref{sec:pyro-model} containing the four sites $({\bf R},i)$, where
${\bf R}$ labels the centers of up-pointing tetrahedra (up-pointing diamond sites). In
this case $S^z_{{\bf R}i} = e_{{\bf R}i}$.

We calculate the equal-time correlator ${\cal C}^E_{ij}({\bf r}-{\bf r}') =
\langle E_i({\bf r}) E_j({\bf r}')\rangle$ in the gauge-fixed continuum
theory for the Coulomb phase,
where $E_i = \epsilon_{ijk}\partial_j \Upsilon_k$. We illustrate the calculation with
a single component:
\begin{eqnarray}
\label{eq:efield-diag}
{\cal C}^E_{zz}({\bf R}) &=& \frac{{\cal K}_c}{2 v_p} \int \frac{d^3{\bf k}}{(2\pi)^3}
\frac{k_x^2 + k_y^2}{k}e^{i {\bf k}\cdot{\bf R}} \\
&=& \frac{{\cal K}_c}{2 v_p} \int_0^{\Lambda} \frac{dk\, k^3}{(2\pi)^3} 
\int d\Omega' \operatorname{sin}^2 \!\theta'
\operatorname{exp}(i k R \operatorname{cos} \gamma) \nonumber\text{,}
\end{eqnarray}
where we have imposed the hard momentum cutoff $\Lambda$, $\gamma$ is the angle between
${\bf k}$ and ${\bf R}$, $\theta'$ is the angle between ${\bf k}$ and the $z$-axis, and
the $\Omega'$ integral is over the angular direction of ${\bf k}$.
Expanding the exponential in Legendre polynomials of
$\operatorname{cos}\gamma$, and using the addition theorem to rewrite these in terms
of spherical harmonics, the integral can be done to find:
\begin{equation}
\label{eq:efield-diag-result}
{\cal C}^E_{zz}({\bf R}) = \frac{{\cal K}_c}{\pi^2 v_p R^4} (2 \operatorname{cos}^2 \!\theta - 1)
\text{.}
\end{equation}
Here $\theta$ is the angle between ${\bf R}$ and the $z$-axis.
We have dropped terms oscillating at the cutoff
wavevector, which are unphysical artifacts of the hard cutoff. Alternatively, we can
use a soft cutoff by inserting $e^{-k/\Lambda}$ in the integrand, integrating over all
$k$-space, taking the limit $R \gg \Lambda^{-1}$ and keeping only the
dominant powers of $R$ to recover the same result.
In the general case we find:
\begin{equation}
\label{eq:efield-general}
{\cal C}^E_{ij}({\bf R}) = \frac{{\cal K}_c}{\pi^2 v_p R^6} (2 R_i R_j - R^2 \delta_{ij})\text{.}
\end{equation}

The striking angular dependence of this correlator is a manifestation of
the inherent vectorial structure of the $U(1)$ spin liquid, which comes
in turn from the vector fields of $U(1)$ gauge theory. For the pyrochlore
model in the $U(1)$ spin liquid phase we have:
\begin{equation}
\label{eq:pyro-spin-spin}
\langle S^z_{{\bf R}i} S^z_{{\bf R}'j}\rangle 
\sim ({\bf e}_i)_k ({\bf e}_j)_l C^E_{kl}({\bf R} - {\bf R}')\text{,}
\end{equation}
where ${\bf e}_i$ are unit vectors connecting ${\bf R}$ with its 
nearest-neighbor
\emph{diamond} sites. In the cubic case:
\begin{equation}
\label{eq:cubic-spin-spin}
\langle S^z_{{\bf r}i} S^z_{{\bf r}'j}\rangle
\sim \operatorname{exp}\big(i {\bf K}_0 \cdot ({\bf r}-{\bf r}')\big)
C^E_{ij}({\bf r} - {\bf r}')\text{.}
\end{equation}

In the RK point effective theory, a similar calculation shows that both the radial and
angular dependence of the correlator changes. It takes the familiar ``dipole'' form:
\begin{equation}
\label{eq:rk-efield-general}
{\cal C}^{E-RK}_{ij}({\bf R}) =
\frac{{\cal K}_c m}{8\pi R^5}(3 R_i R_j - R^2 \delta_{ij}) \text{,}
\end{equation}
falling off as $1/R^3$.

It is also interesting to calculate the correlations of the kinetic energy
density. In our ring exchange models this naturally lives on the plaquettes
and in the cubic case has the form 
\begin{equation}
\label{eq:ke-density}
\varepsilon_{{\sf r}{\sf r}'} = \frac{1}{2}(S^+_1 S^-_2 S^+_3 S^-_4 + h.c.)
= \operatorname{cos}b_{{\sf r}{\sf r}'} \text{.}
\end{equation}
In the pyrochlore model Eq.~(\ref{eq:ke-density}) looks the same, but with
ring exchange on the hexagonal plaquettes as in Eq.~(\ref{eq:pyro-heff}).
Long range order at ${\bf k} \neq 0$ 
in $\langle\varepsilon_{{\sf r}i} \varepsilon_{{\sf r}'j}\rangle$
would indicate a plaquette density
wave state with broken translation symmetry. To simplify notation we
work out the cubic case, where
\begin{eqnarray}
\label{eq:cubic-ke-correlator-1}
\langle \varepsilon_{{\sf r}i} \varepsilon_{{\sf r}'j} \rangle &\sim&
\langle \operatorname{cos}(\tilde{b}_{{\sf r}i}) 
\operatorname{cos}(\tilde{b}_{{\sf r}'j})\rangle_0 \\
&=& \operatorname{exp}(-\langle \tilde{b}^2 \rangle_0)
\operatorname{cosh} (\langle \tilde{b}_{{\sf r}i} \tilde{b}_{{\sf r}'j} \rangle_0) 
\nonumber\text{.}
\end{eqnarray}
The prefactor involving $\langle \tilde{b}^2 \rangle_0$ is a nonuniversal constant.
At large separation, the second factor can be evaluated in the continuum theory, giving:
\begin{equation}
\label{eq:cubic-ke-correlator-2}
\langle \varepsilon_{{\sf r}i} \varepsilon_{{\sf r}'j} \rangle \sim
C\Big[ 1 + \big({\cal C}^B_{ij}({\sf r} - {\sf r}')\big)^2 + \cdots\Big]\text{.}
\end{equation}
We can immediately write down the magnetic field correlator by duality,
which simply interchanges ${\cal U}_c$ and ${\cal K}_c$ within
the Coulomb phase. Therefore
\begin{equation}
\label{eq:bfield-general}
{\cal C}^B_{ij}({\bf R}) = \frac{{\cal U}_c}{{\cal K}_c} {\cal C}^E_{ij}({\bf R})\text{,}
\end{equation}
and the kinetic energy density exhibits power-law $1/R^8$ correlations and a
nontrivial angular dependence.

The RK point theory does not have this self-duality property, so we need to
evaluate another integral to find the magnetic field correlator. Using a hard cutoff,
one finds \emph{only} unphysical oscillatory terms. Using the $e^{-k/\Lambda}$ soft cutoff,
the result is ${\cal C}^{B-RK}_{ij}({\bf R}) \propto
1/\Lambda R^6$. This cutoff-dependence indicates
that the angular dependence is probably nonuniversal, since there should
be other $1/R^6$ contributions with nontrivial angular factors that have been left
out of the continuum theory.

\subsection{Monopole Propagator}
\label{sec:monopole-prop}

In this section we calculate the monopole propagator and verify that it falls off
exponentially in space and imaginary time, both in the Coulomb phase and at the RK point.
This is one of the key properties of the $U(1)$ spin liquid, since at the
transition to any nearby confining phase the monopoles will proliferate and condense. It is most
convenient focus on the equal-time monopole propagator
$\langle m^{\dagger}_{{\sf R}} m_{{\sf R}'} \rangle$,
working on the spacetime lattice.
For ${\sf R} \neq {\sf R}'$,
this expectation value \emph{vanishes} taken with respect to ${\cal S}^0_{lat}$ because it
creates magnetic charge at two points. To understand the true behavior we need to include
magnetic charge fluctuations and add the corrections Eq.~(\ref{eq:non-ginvt-corrections}) to
the Gaussian action.

For simplicity, we consider only the restricted set of correction terms:
\begin{equation}
\label{eq:mag-charge-hopping}
-\delta {\cal S} = 
v^{\tau}_2 \sum_{\tau,{\sf r}} \operatorname{cos}(2 \tilde{\alpha}_{{\sf r}\tau}) +
v_2 \sum_{\tau,\langle{\sf r}{\sf r}'\rangle} \operatorname{cos}(\alpha^0_{{\sf r}{\sf r}'})
\operatorname{cos} (2\tilde{\alpha}_{{\sf r}{\sf r}'}) \text{.}
\end{equation}
We want to calculate the propagator in perturbation theory in $\delta{\cal S}$:
\begin{eqnarray}
\label{eq:mono-prop-calc-1}
\langle \tilde{m}^{\dagger}_{{\sf R}} \tilde{m}_{{\sf R}'} \rangle &=&
\frac{1}{{\cal Z}}   
\prod_{\tau,{\sf r}, \mu}
\int [d\tilde{\alpha}_{{\sf r},\mu}(\tau)] \Bigg[\\
&&\operatorname{exp}(i \sum_{\langle{\sf r}{\sf r}'\rangle} {\cal B}_{{\sf r}{\sf r}'}
\tilde{\alpha}_{{\sf r}{\sf r}'})
\operatorname{exp}(- {\cal S}^0_{lat} - \delta{\cal S})\Bigg] \text{,}\nonumber
\end{eqnarray}
where ${\cal B}$ is the classical magnetic field due to a monopole-antimonopole pair at
${\sf R}$ and ${\sf R}'$, respectively. To obtain a nonvanishing contribution, we need to
bring down correction terms until the complex exponential has been modified to create
no magnetic charges. To lowest order we obtain:
\begin{equation}
\label{eq:mono-prop-calc-2}
\langle m^{\dagger}_{{\sf R}} m_{{\sf R}'}\rangle  \sim
v_2^{|{\sf R}'-{\sf R}|} \Big\langle
\operatorname{exp}
(i \sum_{\langle{\sf r}{\sf r}'\rangle} 
{\cal B}_{{\sf r}{\sf r}'}\tilde{\alpha}_{{\sf r}{\sf r}'})
\prod_{{\sf r}{\sf r}' = {\sf R}}^{{\sf R}'} \!\!\!\!\!\!\!\!\!\!\!\!\longrightarrow
\operatorname{cos}(2\tilde{\alpha}_{{\sf r}{\sf r}'})
\Big\rangle_0 \text{,}
\end{equation}
where the product of cosines is taken over the shortest path connecting ${\sf R}$ and
${\sf R}'$; for simplicity we restrict our attention to geometries where this is unique,
although this is inessential. We have also dropped the background field $\alpha^0$, which
gives only an overall sign. 
The gauge-invariant part of the expectation value comes from the term in the product of 
cosines that threads one magnetic flux quantum
from ${\sf R}$ to ${\sf R}'$, giving a new magnetic field ${\cal B}'$ with zero divergence,
and allowing us to express the result as the expectation value of a gauge-invariant operator:
\begin{equation}
\label{eq:mono-prop-calc-3}
\langle m^{\dagger}_{{\sf R}} m_{{\sf R}'}\rangle  \sim
\Big(\frac{v_2}{2}\Big)^{|{\sf R} - {\sf R}'|} 
\langle \operatorname{exp}(i \sum_{\langle{\sf r}{\sf r}'\rangle}{\cal B}'_{{\sf r}{\sf r}'}
\tilde{\alpha}_{{\sf r}{\sf r}'}) \rangle_0 \text{.}
\end{equation}
What the perturbation theory has done is exactly to connect the monopole and antimonopole
with a Dirac string.

The exponential decay is clear from the prefactor in Eq.~(\ref{eq:mono-prop-calc-3}). To
evaluate the corrections to this, 
we apply the Faddeev-Popov procedure to the functional integral for the gauge-invariant
expectation value,
and calculate using the photon propagator. We could \emph{not} have done
this at the outset since the original operator was not gauge-invariant. It is most convenient
to integrate by parts inside the exponential to obtain a result in terms of the classical 
vector potential and the electric field:
\begin{equation}
\label{eq:dirac-string-mono-prop}
\langle \operatorname{exp}(i \sum_{\langle{\bf r}{\bf r}'\rangle} {\cal A}'_{{\bf r}{\bf r}'}
\tilde{e}_{{\bf r}{\bf r}'} ) \rangle_0 =
\operatorname{exp}\Big(-\frac{1}{2} \sum_{{\bf r},{\bf r}',i,j}
{\cal A}'_{{\bf r}i} {\cal A}'_{{\bf r}'j} \langle \tilde{e}_{{\bf r}i}
\tilde{e}_{{\bf r}' j} \rangle_0 \Big) \text{.}
\end{equation}
The necessary integral is quite difficult to evaluate, so we resort to power-counting
to determine the largest possible contribution. Consider a spatial separation of $R$ between
monopole and antimonopole. One contribution will come from the region near the pair, 
giving a factor of $R^6$ from the integration, $R^{-2}$ from the two vector potential factors,
and $R^{-4}$ from the electric field correlator in the Coulomb phase. These factors multiply
to give a constant, so we expect that the largest possible contribution to the integral is
logarithmic in $R$, which contributes only a power-law prefactor to the propagator. Other
contributions involving regions far away from the pair make subdominant contributions. At the RK
point there is an extra power of $R$ from the electric field correlator, and the dominant
possible contribution is linear in $R$ and gives a correction to the correlation length.

Finally it is clear from these considerations that the unequal-time propagator
is not substantially different, and decays exponentially in space and time.

\section{Exact Ground State Wavefunction}
\label{sec:rkpoint}

In this section we return to the microscopic spin models, and use the exact ground state
at the Rokhsar-Kivelson point to extract information about the physics nearby. We begin with a
discussion of the structure of the wavefunction and some simple properties that can be seen
analytically, then proceed to a discussion of the numerical evaluation of several quantities.
In addition to various equal time correlation functions, we make use of a remarkable property
of RK-type points discovered by Henley\cite{henley} to approximately evaluate the imaginary-time
monopole propagator.

We will consider finite-size cubic and pyrochlore lattices in this section. To fix notation,
for the cubic lattice $L$ denotes the number of sites on a side of a periodic system. In the
pyrochlore case we take periodic boundary conditons so that ${\bf R}$ and ${\bf R} + L{\bf a}_i$
are identified, and we can think of $L$ as the number of tetrahedra along one periodic direction.

\subsection{General Properties}
\label{sec:rk-general-props}

The RK wavefunction is an equal-weight superposition (with positive coefficients)
of all possible dimer coverings consistent with the local dimer constraint. In the spin
models the electric field can only take on the values $\pm 1/2$, and it is convenient to
visualize dimer coverings as zero-divergence configurations of an electric field that can
point only forward or backward on each link. Within the vast superposition there are components
of every electric topological sector. Since these are not mixed by the dynamics they should
be thought of as degenerate but distinct ground states. The RK point
exhibits the electric sectors of $U(1)$ topological order, and as in the Gaussian RK action
they are exactly degenerate even in a finite-size system.

To understand what sector to focus on, it is profitable to consider first-order
perturbation theory in $\delta V N_f$ away from the RK point. The first-order
shift in the ground state energy is $\delta E = \delta V \langle N_f\rangle$,
where $\langle N_f\rangle$ is the average
number of flippable plaquettes \emph{in a given sector}. For $\delta V > 0$, sectors with
\emph{no} flippable plaquettes, and hence no dynamics, will have the lowest energy, and we reach
the usual conclusion that a valence bond crystal obtains on this side of the RK point.
For $\delta V < 0$, however, the sector with the greatest average flippability wins out. This
is in some sense the most disordered sector, and should have $\Phi^E_i = 0$; several numerical
checks support this conclusion. We are interested in $\delta V < 0$, so we focus on the
zero-flux sector.

We also wish to consider sectors that violate the zero-divergence constraint and
contain spinons. Within each such sector, the lowest energy state is again given by an
equal-weight superposition of all electric field configurations consistent with a
background charge density specifying the spinon positions. These are exact eigenstates
with energy $J_z/2$ times the number of spinons, which in a finite-size system is always even.
Consider in particular a sector with two spinons. There is clearly zero energy cost no matter
how the spinons are moved around, so the RK point is deconfined and has no Coulomb potential
between static electric charges. Again, this is consistent with the effective action
Eq.~(\ref{eq:rk-effective-action}).

Finally, we calculate the variational energy of the magnetic topological sectors. In general,
we create a classical configuration of the vector potential by acting on the RK ground
state with the operator
${\cal O} = \operatorname{exp}(i \sum_{\langle{\bf r}{\bf r}'\rangle} {\cal A}_{{\bf r}{\bf r}'}
e_{{\bf r}{\bf r}'})$, just as for the special case of monopole creation in
Eq.(\ref{eq:micro-monopole-creation}). Of course we expect only that the resulting state
has some overlap with the desired eigenstate of the gauge theory. Denote the RK wavefunction
by
\begin{equation}
|\psi_{RK}\rangle =
\frac{1}{\sqrt{{\cal N}}} \sum_{\{e_{{\bf r}i}\}} |\{ e_{{\bf r}i} \}\rangle \text{,}
\end{equation}
where the sum is over all configurations of the electric field in the desired sector and
${\cal N}$ is the number of states contributing to the sum. Defining $N_f[\{e\}]$ to be
the number of flippable plaquettes in a given configuration, and $P_f[\{e\},\square]$ to
be one if the specified plaquette is flippable and zero otherwise, 
the variational energy of interest is given by:
\begin{eqnarray}
E_{var}[{\cal A}] &=& \langle \psi_{RK} | {\cal O}^{\dagger} {\cal H}_{RK}\,
{\cal O} | \psi_{RK} \rangle \\
&=& \frac{J_{ring}}{{\cal N}} \sum_{\{e\}} \Big[ N_f[\{e\}] \nonumber \\
&&- \sum_{\square} P_f[\{e\},\square] 
\operatorname{exp}(2 i \sum_{{\bf r}{\bf r}' \in \square} {\cal A}_{{\bf r}{\bf r}'}
e_{{\bf r}{\bf r}'}) \Big] \nonumber\\
&=& \frac{J_{ring}}{{\cal N}}
\sum_{\{e\},\square} P_f[\{e\},\square] 
\Big[1 - \operatorname{cos}\big((\operatorname{curl} {\cal A})_{\square} \big) \Big]\text{.}
\nonumber
\end{eqnarray}
Here we use notation appropriate to the cubic model; the generalization to the pyrochlore
is obvious.

Specifically, consider the case where ${\cal O}$ threads one quantum of magnetic flux
through the $x$-direction in the cubic model. Then $\operatorname{curl}{\cal A} = 2\pi/L^2$
for plaquettes in the $x$-direction and zero otherwise, and
\begin{equation}
E_{var} = n_f L^3 J_{ring}\big(1 - \operatorname{cos}(2\pi/L^2)\big)\text{,}
\end{equation}
where $n_f$ is the average flippability per plaquette. For $L \to \infty$ this energy goes to
zero as $1/L$ as expected, and further gives a rough value for the ``magnetic stiffness''
at the RK point: ${\cal K} \approx n_f J_{ring}$. Using the numerical methods discussed below, 
for the cubic model in the zero-flux sector 
we find $n_f \approx 0.260$. For the pyrochlore a similar calculation
shows ${\cal K} \approx 2 n_f J_{ring}$; the factor of two arises because the flux passes
through two kinds of hexagons. In this case we find numerically $n_f \approx 0.175$.

\subsection{Monte Carlo Algorithms and Ergodicity}

Because the RK wavefunction has positive and equal weights, equal-time properties can be
evaluated by infinite temperature Monte Carlo for the associated \emph{classical} dimer model.
The simplest possibile Monte Carlo step is to: 1) Randomly choose a plaquette. 2) If the
plaquette is flippable, flip it, otherwise do nothing. As desired,
this algorithm preserves the electric flux $\Phi^E_i$. For the measurement of the quantities
discussed below, the algorithm was run for as many as $10^{11}$ Monte Carlo steps at a given
system size. In some cases these long runs were necessary to achieve good accuracy because
the desired quantity was very small.

It is not clear whether the single ring move algorithm is ergodic within
each electric field sector.
For $L=2$ cubic and pyrochlore systems we have performed an exact enumeration of all allowed
configurations. In the cubic case the zero-flux sector contains 880 states; 864 of these have
flippable plaquettes (with $n_f = 1/3$ on average), and are connected under single ring
moves. The other 16 states have no flippable plaquettes. For the pyrochlore model there are
384 zero-flux states connected under single ring moves ($n_f$ = 1/4), and 12 zero-flux states
with no flippable hexagons.
Evidently the single ring move algorithm is not ergodic within the zero-flux sector, but it does
not matter if we fail to access states with no flippable plaquettes. In our simulations 
we generate
initial configurations for larger lattices by periodically repeating a state from the flippable
part of the $L=2$ zero-flux sector, so more insidious problems could occur if this sector
breaks up into multiple flippable subsectors in larger lattices, each closed under single ring
moves.

\begin{figure}
\begin{center}
\centerline{\includegraphics[width=3.3in]{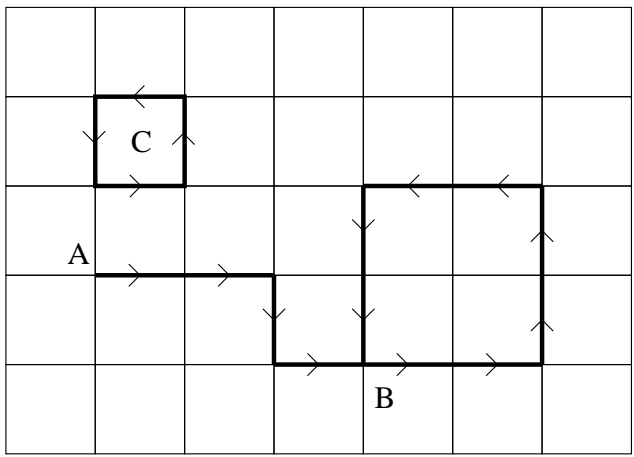}}
\caption{Illustration of the loop move algorithm of Barkema and Newman looking at one plane
of the cubic model. One randomly
chooses a starting point (A), and executes a random walk along the lattice bonds,
moving only along the direction of the electric field vectors. Once the random walk
intersects itself (B), the tail is removed and the electric field is reversed everywhere along
the resulting loop, preserving the zero-divergence constraint. If only contractible loops are
desired (as in our simulations within the zero-flux sector), one can simply throw away
non-contractible loops and repeat these steps until a contractible loop results; this
does not affect detailed balance. The single ring move only reverses the electric fields
around elementary plaquettes like that at (C).}
\label{fig:ice-algorithm}
\end{center}
\end{figure}

For an analogous two-dimensional model, it is possible to prove that single ring moves are
ergodic in each electric flux sector (see Appendix \ref{app:2d-ergodic}). In the absence of
a similar result 
in three dimensions, we have checked some of our results for the cubic model with an
algorithm due to Barkema and Newman that we believe is probably ergodic\cite{ice-algorithm}.
This algorithm is illustrated in Figure \ref{fig:ice-algorithm}; in the present case
the basic idea is to flip loops of arbitrary length (including single rings), keeping
only non-contractible loops that do not change $\Phi^E_i$. All properties measured using both
algorithms gave the same results; we present two examples in the following section. While
we have not implemented an algorithm with ``loop moves'' 
for the pyrochlore model, due to the great similarity of all measured properties 
to their cubic analogs it seems unlikely that ergodicity is an issue.

\subsection{Equal Time Properties}

\begin{figure}
\begin{center}
\vskip+6mm
\centerline{\includegraphics[width=3.3in]{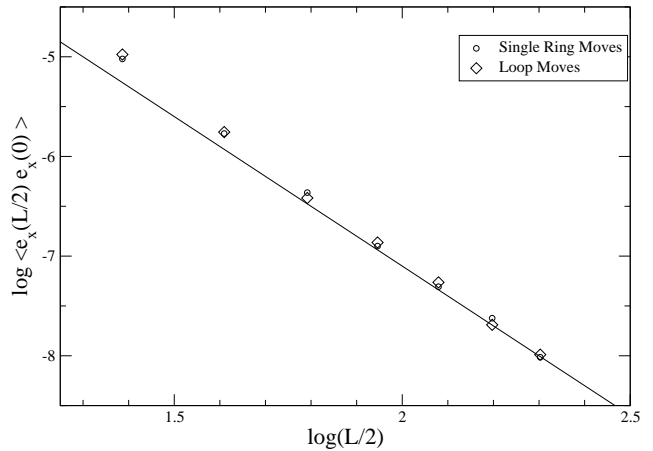}}
\caption{Log-log plot of the equal time electric field correlator at the RK point of 
the cubic model, in the orientation discussed in the text. The circles denote data
from the single ring move algorithm, while the diamonds were obtained by the loop
move algorithm. Error bars are on the order of the symbol size. The line is a guide to
the eye with slope $-3$ to show the very good agreement with the $1/R^3$ decay expected
from the effective action.}
\vskip-4mm
\label{fig:cubic-ee}
\end{center}
\end{figure}

\begin{figure}
\begin{center}
\vskip+6mm
\centerline{\includegraphics[width=3.3in]{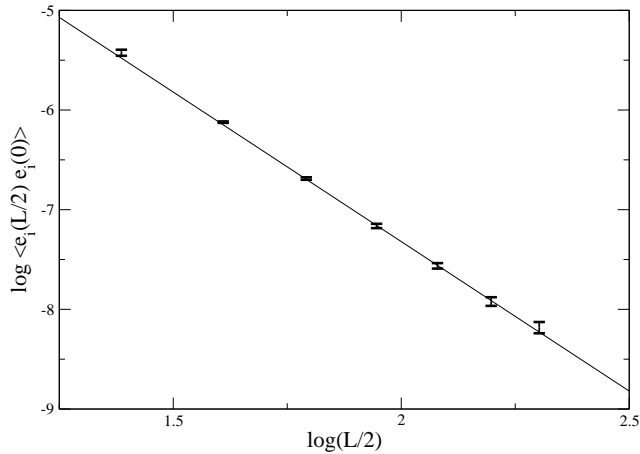}}
\caption{Log-log plot of the pyrochlore RK point electric field correlator
in the orientation discussed in the text. As in Fig. \ref{fig:cubic-ee}, the line is a guide to
the eye with slope $-3$.}
\vskip-4mm
\label{fig:pyro-ee}
\end{center}
\end{figure}

We measured the electric field-electric field correlator in both the cubic and pyrochlore
models. To extract the asymptotic dependence most simply we focused on the correlators
measured at a separation of half the system size. It should be noted that one cannot trivially
extract the angular dependence discussed in Section \ref{sec:power-laws} by this method, since
the two electric field vectors in the correlator will be connected by many paths with length
scaling as $L$ but with different angles. For the cubic model we
show our results for $\langle e_x(L{\bf x}/2) e_x({\bf 0}) \rangle$ in Figure \ref{fig:cubic-ee}.
Data from both algorithms are shown and the agreement with the $1/R^3$ decay predicted by
the effective theory is very good. The same conclusion obtains for the pyrochlore model
(but with only the single ring move algorithm). In that case the correlator measured was
$\langle e_0(L{\bf a}_1/2) e_0({\bf 0}) \rangle$; the data are shown in Figure \ref{fig:pyro-ee}.
Measurements of other orientations of the electric field correlator in both models all showed
the same $1/R^3$ decay.

\begin{figure}
\begin{center}
\vskip+6mm
\centerline{\includegraphics[width=3.3in]{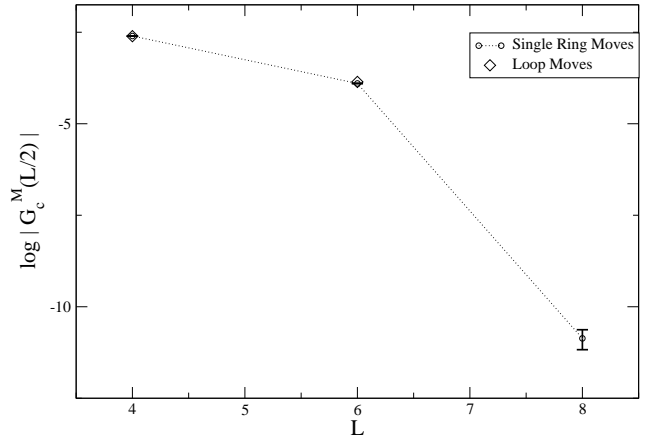}}
\caption{Cubic model equal-time monopole propagator. The vertical axis is the logarithm
of the absolute value of 
the double-strength propagator discussed in the text, and the horizontal axis is
system size. Data for both the single ring move (circles) and loop move (diamonds) algorithms
are shown. For the loop move data the error bars are smaller than the symbol size, and
for the first two single ring move data points they are obscured by the other symbols.
The results are consistent with exponential decay with power-law corrections.}
\vskip-4mm
\label{fig:cubic-mono}
\end{center}
\end{figure}

To measure the equal time monopole propagator it is necessary to address some subtleties that
arise when putting objects with gauge charge in a finite-size system. First, it is clear
that because the system is a compact manifold with no boundary,
it must have zero total magnetic charge. Since the operator
$m^{\dagger}_{{\sf R}} m_{{\sf R}'}$ creates a monopole-antimonopole pair, this is not a problem
for the propagator. However, consider the cubic lattice and let
${\sf R}-{\sf R}' = n_x{\bf x} \neq 0$. For any value of $n_x$ there will be planes where
$\Phi^B_x \neq 2\pi n^B_x$, which is not allowed for a compact vector potential as discussed
in Section \ref{sec:toporder}. We can ameliorate this problem by creating a \emph{double strength}
monopole-antimonopole pair, but even in this case the only allowed separation is $n_x = L/2$.
Because of these complications, in the cubic case we only measure the double-strength propagator
at a separation of half the system size:
\begin{equation}
\label{eq:finite-size-cubic-mono-prop}
G^M_c(L/2) = \langle \big(m^{\dagger}_{({\sf R}+L{\bf x}/2)}\big)^2 \big(m_{{\sf R}}\big)^2 \rangle
\text{.}
\end{equation}
By time-reversal or Ising symmetry, $G^M_c$ is real. We obtained the vector potential appropriate
for Eq.~(\ref{eq:finite-size-cubic-mono-prop}) by first solving a discrete Poisson's equation
by numerical matrix inversion, then feeding the solution into 
Eq.~(\ref{eq:monopole-classical-energy}).
This was minimized by a combination of simulated annealing
and direct minimization (i.e. zero-temperature simulated annealing).
Because of the very rapid decay, it was only practical to measure $G^M_c$ for $L \leq 8$;
results are shown in Figure \ref{fig:cubic-mono} and are consistent with exponential decay
with power-law corrections.

\begin{figure}
\begin{center}
\vskip+4mm
\centerline{\includegraphics[width=3.3in]{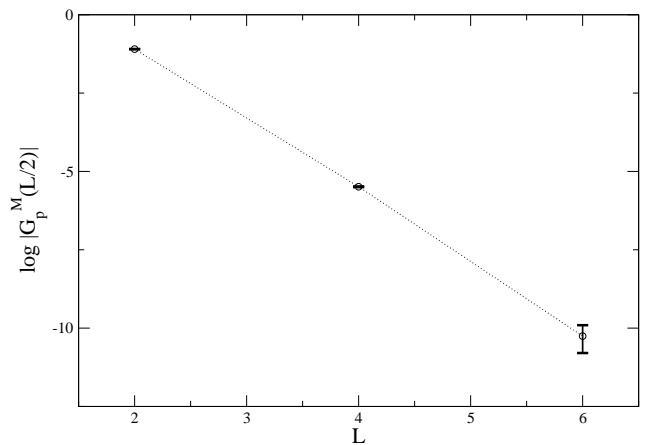}}
\caption{Pyrochlore equal-time monopole propagator. The vertical axis is the logarithm
of the absolute value of the propagator Eq.~(\ref{eq:finite-size-pyro-mono-prop}). The results
are consistent with exponential decay, with no indication of power-law corrections.}
\label{fig:pyro-mono}
\end{center}
\end{figure}

On the pyrochlore lattice it is not possible to have
double-strength magnetic charges in the microscopic model 
(see Sec. \ref{sec:pyro-mappings}), so for consistency with
flux quantization we must consider a more complicated geometry. We look at the following
propagator, which creates two monopoles on adjacent up- and down-pointing (dual) diamond sites, 
and similarly two antimonopoles separated by a distance $L/2$:
\begin{eqnarray}
\label{eq:finite-size-pyro-mono-prop}
G^M_p(\!\!\!&L&\!\!\!\!/2) = \\ 
&\langle& \!\!\!\!m^{\dagger}_{({\sf R}+L{\bf a}_0/2)} m^{\dagger}_{({\sf R}'+ L{\bf a}_0/2)}
m_{{\sf R}} m_{{\sf R}'} \rangle \nonumber\text{,}
\end{eqnarray}
where ${\sf R}$ is an up-pointing diamond site and ${\sf R}'$ is the down-pointing
site directly above. The results are shown in Figure \ref{fig:pyro-mono}, and are consistent
with exponential decay; we do not believe the apparent lack of substantial power-law
corrections is significant, since the coefficient of these corrections is presumably
nonuniversal.

\begin{figure}
\begin{center}
\centerline{\includegraphics[width=3.3in]{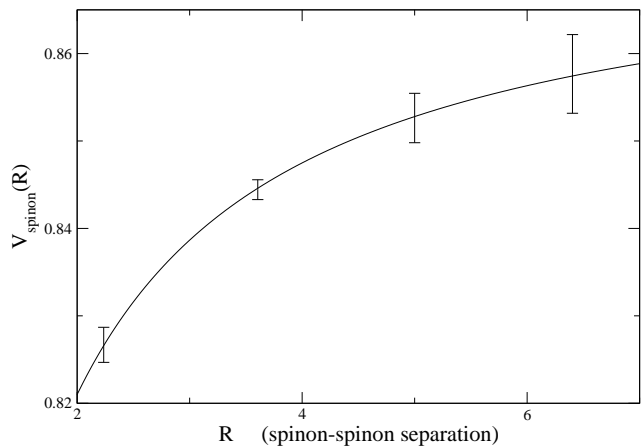}}
\caption{Plot of the spinon-spinon potential defined in Eq.~(\ref{eq:spinon-potential}) for
several system sizes of the cubic model. The horizontal axis is the 
distance between the spinons, $R = \sqrt{L^2/2 - L + 1}$.
The curve is a fit of the data to the functional form $V_1 + V_2/R$.}
\label{fig:cubic-spinon-potential}
\vskip-4mm
\end{center}
\end{figure}

In the cubic model, we also measured the potential between a pair of static spinons in
first-order perturbation theory away from the RK point. In the notation of 
Sec.~\ref{sec:rk-general-props}, we consider $\delta V < 0$. The coefficient of $|\delta V|$ in
the spinon-spinon potential is given by:
\begin{equation}
\label{eq:spinon-potential}
V_{spinon}({\bf R}) = \big\langle N_f(\text{no spinons}) - N_f(\text{spinons})\big\rangle\text{.}
\end{equation}
To measure this quantity, we generated sectors with one $S^z = 1/2$ spinon at ${\bf r} = 0$,
and another at ${\bf r} = (L/2){\bf x} + (L/2 - 1){\bf y}$ for $L \geq 4$. The results are shown
in Figure \ref{fig:cubic-spinon-potential}, and are consistent with a $1/r$ Coulomb potential.
While the perturbation theory does not necessarily ``know'' whether a deconfining phase with a
Coulomb potential can be stable, it does indicate ${\cal U}_c > 0$ for
small $\delta V < 0$. Since the Coulomb phase \emph{is} stable, we conclude it indeed
exists adjacent to the RK point over a finite range $0 > \delta V > \delta V_{min}$.

\subsection{Monopole Gap}

In our models 
it is possible to approximately measure quantum imaginary time correlation functions using
only the classical Monte Carlo dynamics. The key observation, due
to Henley\cite{henley}, is that at the RK point the master equation describing the Monte Carlo
dynamics of the single ring move algorithm is \emph{identical} to the imaginary-time
Schr\"{o}dinger equation. We will be interested in monopole
correlation functions built from the equal-time correlators constructed above.
In the cubic case, for example, we measure:
\begin{equation}
\label{eq:cubic-imag-mono}
C^M_c(\tau) = \langle \hat{{\cal O}}_M
\operatorname{exp}(-\tau {\cal H}_{RK})
\hat{{\cal O}}^{\dagger}_M \rangle \text{,}
\end{equation}
where 
$\hat{{\cal O}}^{\dagger}_M = \big(m^{\dagger}_{({\sf R}+L{\bf x}/2)}\big)^2
\big(m_{{\sf R}}\big)^2$.
The analogous definition of $C^M_p(\tau)$ for the pyrochlore model is constructed from
the equal-time propagator Eq.~(\ref{eq:finite-size-pyro-mono-prop}). 

To extract information about the spectrum
from the simulation, it is useful to relate the units of classical Monte Carlo time
to quantum imaginary time. Let $N_p$ be the total number of plaquettes in the system. If we begin
with some electric field configuration $\{e\}$ and execute one Monte Carlo step,
the probability of remaining in the same state is $(1 - N_f[\{e\}]/N_p)$, while that of making
a transition to each of the states accessible by flipping one plaquette is $1/N_p$. We now
consider the time-discretized imaginary-time Schr\"{o}dinger equation and fix the timestep
$\Delta\tau$ to recover the same values, which now enter as probabillity \emph{amplitudes}.
We consider:
\begin{eqnarray}
\label{eq:imag-time-schrodinger}
\operatorname{exp}\big(-\!\!\!\!&\Delta&\!\!\!\!\tau\,{\cal H}_{RK}\big)|\{e\}\rangle \\
&=& \Big(1 - \Delta\tau\,{\cal H}_{RK} +
{\cal O}\big((\Delta\tau{\cal H}_{RK})^2\big)\Big)|\{e\}\rangle \nonumber\\
&=& \big(1 - J_{ring}\Delta\tau N_f \big)|\{e\}\rangle
+ J_{ring}\Delta\tau \sum_{\{e'\}} |\{e'\}\rangle
\text{,}\nonumber
\end{eqnarray}
where the final sum is over those electric field configurations connected to $\{e\}$ by a
single ring move. It is only valid to neglect the higher-order terms in $\Delta\tau{\cal H}_{RK}$
when $N_f/N_p \ll 1$; in fact, the correspondence between the \emph{discrete}-time quantum
dynamics and the Monte Carlo dynamics (which are necessarily discrete in our simulation), is
only strictly valid in this limit. In our models $N_f/N_p$ is of order $1/5$, so we expect to
at best make quantitative errors of about $10\%$, and at worst to get the wrong answer. This
problem cannot be alleviated by the means at hand because the discrete
classical algorithm \emph{fixes} the time-step for the quantum dynamics. A more careful treatment
would require a direct simulation of the master equation with control over
$\Delta\tau$, or quantum Monte Carlo.

If we nevertheless expand the exponential,
setting $\Delta\tau = (J_{ring}N_p)^{-1}$ correctly matches the classical 
probabilities and quantum amplitudes.
Using this relation, we write down the appropriate quantity to simulate to measure the
imaginary time correlator:
\begin{eqnarray}
C^M(\tau &=& t/J_{ring}N_p) 
= \frac{1}{{\cal N}}\sum_{\{e\}} \frac{1}{N_{{\cal T}(t,\{e\})}}\Big[ \\
&&\!\!\!\!\!\!\!\!
\sum_{{\cal T}(t,\{e\})} \big({\cal O}^M(\{e\})\big)^* {\cal O}^M\big({\cal T}(t,\{e\})\big)\Big]
\nonumber\text{.}
\end{eqnarray}
Here $t$ is an integer number of steps in Monte Carlo time, and ${\cal T}(t,\{e\})$ labels
all possible Monte Carlo time evolutions of time $t$ starting from the configuration $\{e\}$.
The number of such evolutions is $N_{{\cal T}(t,\{e\})}$, and ${\cal N}$ is the total number
of electric field configurations in the zero-flux sector. Finally, we have defined
$\hat{{\cal O}}^M|\{e\}\rangle = {\cal O}^M(\{e\})|\{e\}\rangle$.

\begin{figure}
\begin{center}
\vskip+6mm
\centerline{\includegraphics[width=3.3in]{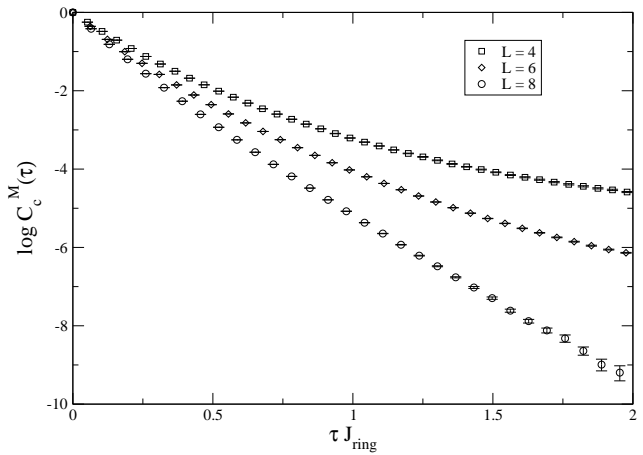}}
\caption{Logarithmic plot of the approximate monopole-antimonopole imaginary time correlator
$C^M_c(\tau)$
for the cubic model. The horizontal axis is imaginary time in units of $J_{ring}^{-1}$.
Data is shown for $L = 4,6,8$; as discussed in the text, as $L$ increases the exponential
decay in imaginary time becomes cleaner, suggesting the monopole is indeed gapped.}
\label{fig:cubic-mgap}
\end{center}
\end{figure}

\begin{figure}
\begin{center}
\centerline{\includegraphics[width=3.3in]{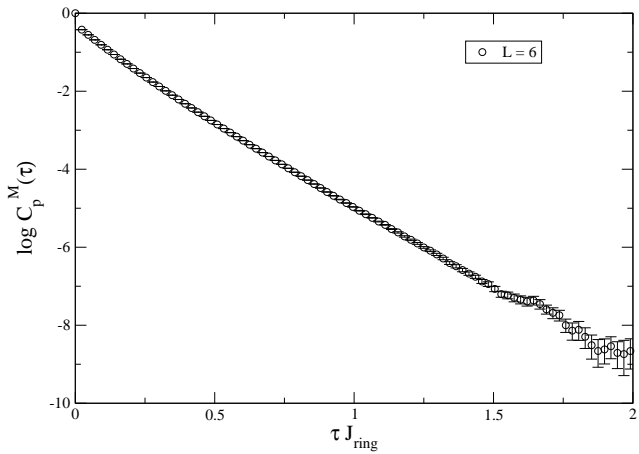}}
\caption{Logarithmic plot of $C^M_p(\tau)$, the approximate monopole-antimonopole imaginary time
correlator in the pyrochlore model. The horizontal axis is imaginary time in units
of $J_{ring}^{-1}$.}
\label{fig:pyro-mgap}
\vskip-4mm
\end{center}
\end{figure}

We have measured the monopole imaginary time correlators for both the cubic and pyrochlore
models; results are shown in Figures \ref{fig:cubic-mgap} and \ref{fig:pyro-mgap}. Note that
the extreme long time behavior is not related to the gap; instead, $\hat{{\cal O}}^M |\psi_{RK}\rangle$
will have
some overlap with the ground state, causing $C^M(\tau)$ to approach a constant as
$\tau \to \infty$. As we go to larger system sizes and the \emph{spatial} separation of
the monopole-antimonopole pair in the correlator increases, there is a clean
exponential decay persisting for longer times. This is illustrated for
several system sizes of the cubic model in Fig.~\ref{fig:cubic-mgap}. From the numerical
data one can extract a rough value for the energy of the monopole-antimonopole configuration
put in at $\tau = 0$; in both cases this is about $4.5 J_{ring}$.
While these results are approximate, it is unlikely that the
monopoles are somehow gapless given the circumstantial evidence for a gap, 
particularly the very rapid exponential decay of the equal-time propagator.

\section{Discussion}
\label{sec:discussion}

In this paper we have argued for the existence of a novel
fractionalized quantum disordered state, the $U(1)$ spin liquid, in
spin models with a global $U(1)$ symmetry. While we found it necessary
to add an extra interaction to the easy-axis limit of the pyrochlore
Heisenberg model to proceed analytically, we speculate that this may
be only a crutch and that the pure easy-axis model is in the
$U(1)$ spin liquid phase.  Both the pyrochlore and cubic models have
no sign problem, so this issue could be resolved by quantum Monte
Carlo simulations\cite{numerical-remark}.  While it would be a
remarkable result if the easy-axis pyrochlore Heisenberg
antiferromagnet were a $U(1)$ spin liquid, we believe the results in
this paper are enough of an ``existence proof'' that the phase is
likely to be present in less tractable but more realistic microscopic
models.

The converse problem, of determining the general requirements for a
spin model to exhibit an emergent $U(1)$ gauge structure, is of course
more challenging. It would appear that, since monopole excitations
are required to enable a transition to a neighboring confined phase,
any such gauge theory must be compact, and hence appear 
\emph{on the lattice}. Thus, unlike emergent global symmetries, which can appear
at low energies and long wavelengths, an emergent $U(1)$ gauge
structure would appear to admit only a limited degree of spatial
coarse-graining. Given the additional need for spin-carrying degrees
of freedom, the unit cell of an emergent gauge theory in a pure spin
model would appear to necessarily contain more than a single $S=1/2$
spin.  Based on the present examples, which take advantage of the
bipartite structure of the cubic and diamond lattices, it is tempting
to speculate that the models should contain a natural bipartite
sublattice. It is easy to see, however, that the emergent gauge
structure in these models is stable to adding arbitrary weak
additional multi-spin interactions, including those that break any
bipartite symmetries. Further understanding, including the important
issue of whether the $U(1)$ spin liquid is possible in a magnet with
global $SU(2)$ symmetry, and if yes, of what its properties might be,
must await further study.

Certainly, the most interesting issue is whether the $U(1)$
spin liquid can be found in a real material. The simplest signature is
the additive $T^3$ contribution to the specific heat from the photon. Because it
is likely to be possible to quantitatively understand the \emph{phonon} specific heat
via independent measurements\cite{broholm}, this could provide a relatively clear
and simple test for emergent photons.
Further theoretical work may be necessary to understand more
delicate probes of the photon, like heat current and (possibly) Raman
scattering; phonon-``photon'' interactions and disorder may play a
nontrivial role. While disorder may be important for some
experimental properties, we remark that simple arguments
demonstrate that disorder does not destabilize the $U(1)$ gauge
structure or the gapless photon.

Since topological order can coexist with conventional broken symmetry,
spin liquids are not the only good places to look for emergent
photons. One can imagine, for example, condensing spin-carrying but
gauge-neutral excitations; in our models this would lead to an ordered
magnetic state with gapless spin waves and a $U(1)$ gauge structure.
Understanding the possibilities for and properties of phases near the
$U(1)$ spin liquid is another problem worthy of consideration if
contact is to be made with experiment. Very recently, Senthil, Vojta
and Sachdev have made the interesting suggestion that $U(1)$-fractionalization
may provide the answer to some of the puzzles of heavy fermion materials\cite{svs}.
We leave these intriguing
issues aside to remark that $U(1)$-fractionalized states are a
remarkable and novel possibility for the physics of strongly
correlated electrons in \emph{three} dimensions, heretofore a
relatively unexplored area.

Finally, we note that Huse, Krauth, Moessner and Sondhi are
considering a model closely related to ours\cite{dimer-3d}.

\begin{acknowledgments}

We thank C. Broholm, C. L. Henley, 
R. G. Melko, R. Moessner, O. I. Motrunich, R. D. Sedgewick, T. Senthil and A. Vishwanath
for numerous stimulating and helpful discussions.
M.H. is grateful for support from the Department of Defense
through the NDSEG program, and from UC Santa Barbara. M.P.A.F. was generously supported by the
NSF under grants DMR-0210790 and PHY-9907949. L.B. was supported by the NSF through
grant DMR-9985255, and by the Sloan and Packard foundations. 
\end{acknowledgments}

\appendix
\section{Enumeration of Symmetries}
\label{app:symmetries}

We here enumerate the action of the various symmetries of the cubic model
on all the microscopic operators, both in dual and original variables. 

\begin{itemize}
\item Discrete translations: ${\bf r} \rightarrow {\bf r} + {\bf 
  R}$. The electric field transforms as $e_{{\bf r}i} \rightarrow e_{{\bf  
  r}+{\bf R},i}$, with similar expressions for $b$ and $a$. Because of the 
  non-translation-invariant background $\alpha^0$, the dual vector 
  potential 
  obeys the more complicated transformation law: 
  \begin{equation}
  \label{eq:dvp-translations}
  \alpha_{{\sf r}i} 
  \rightarrow \alpha_{{\bf R}+{\sf r},i} + \delta\alpha_{{\bf R}+{\sf 
  r},i}
  \text{,}
  \end{equation}
  where the shift $\delta\alpha$ is defined to satisfy
  \begin{equation}
  \frac{1}{\pi} (\operatorname{curl} \delta\alpha)_{{\bf r}+{\bf R},i} =
  e^0_{{\bf r}+{\bf R},i} - e^0_{{\bf r}i}\text{.}
  \end{equation}
\item Lattice rotations: Let ${\cal R}$ rotate the lattice into itself 
  about
  some fixed origin. For example, take the origin to be
  ${\bf r} = 0$ and make a $\pi/2$-rotation about the $z$-axis.
  Then ${\cal R}[n_x {\bf x} + n_y {\bf y} + n_z {\bf z},x] =
  [-n_y {\bf x} + n_x {\bf y} + n_z {\bf z},y]$, and so on.
  We have $e_{{\bf r}i} \rightarrow e_{{\cal R}[{\bf r}i]}$, and 
  similarly for $b$ and $a$. Again $\alpha_{{\sf r}i} \rightarrow 
  \alpha_{{\cal R}[{\sf r}i]} + \delta\alpha_{{\cal R}[{\sf r}i]}$, with
  $(\operatorname{curl} \delta\alpha)_{{\cal R}[{\bf r}i]} = \pi 
  (e^0_{{\cal 
  R}[{\bf r}i]} - e^0_{{\bf r}i})$.
\item Reflections: Let ${\cal F}$ denote reflection about a plane with 
  normal vector ${\bf x}$, ${\bf y}$ or ${\bf z}$. For definiteness, 
  choose the plane with normal ${\bf x}$ at $x = 0$, then:
  \begin{eqnarray}
  e_{{\bf r}x} &\rightarrow& e_{{\cal F}({\bf r}),-x} = -e_{{\cal F}({\bf 
  r})-{\bf x},x} \\
  e_{{\bf r}i} &\rightarrow& e_{{\cal F}({\bf r}),i} \nonumber \text{,}
  \end{eqnarray}
  for $i = y,z$. The vector potential obeys a similar transformation law. 
  The magnetic field and dual vector potential are 
  \emph{pseudovectors} and transform under reflections with an additional 
  minus sign. Again, $\alpha$ transforms with an appropriate shift 
  $\delta\alpha$ to compensate for changes in the background under 
  reflections.
\item Gauge and dual gauge invariance: Under gauge 
transformations, 
  $a_{{\bf r}{\bf r}'} \rightarrow a_{{\bf r}{\bf r}'} + \chi_{{\bf r}'} - 
  \chi_{{\bf r}}$, where $\chi$ is a phase defined on the cubic sites. 
  Similarly we 
  have the dual gauge transformations $\alpha_{{\sf r}{\sf r}'} 
  \rightarrow
  \alpha_{{\sf r}{\sf r}'} + \lambda_{{\sf r}'} - \lambda_{{\sf r}}$, with 
  $\lambda_{{\sf r}} \in \pi {\mathbb Z}$. In the action the dual gauge 
  transformations can be spacetime-dependent, with
  \begin{eqnarray}
  \label{eq:space-gauge-transformations}
  \alpha_{{\sf r}i}(\tau) &\rightarrow& \alpha_{{\sf r}i}(\tau) +  
  \lambda_{{\sf r} + {\bf e}_i}(\tau) - \lambda_{{\sf r}}(\tau) \\
  \alpha_{{\sf r}\tau} &\rightarrow& \alpha_{{\sf r}\tau} + \lambda_{{\sf 
  r}}(\tau + \epsilon) - \lambda_{{\sf r}}(\tau) \nonumber\text{.}
  \end{eqnarray}
\item Ising (or particle-hole) symmetry: $e \rightarrow -e$, $a 
  \rightarrow -a$ and $b \rightarrow -b$. Also $\alpha \rightarrow 
  \alpha^0  - \alpha$.
\item Time reversal: Since spin operators obey
${\bf S} \rightarrow -{\bf S}$ under time-reversal, we have:
$e \rightarrow -e$, $a \rightarrow a + 
  \pi$ and $b \rightarrow b$. For consistency with the electric field 
  transformation law, $\alpha \rightarrow \alpha^0 - \alpha$. In the 
  action 
  the latter relation continues to hold, as long as we also send $\tau 
  \rightarrow -\tau$. Furthermore, the temporal component of the dual 
  vector potential transforms as $\alpha_{{\sf r}\tau}(\tau) \rightarrow
  - \alpha_{{\sf r}\tau}(-\tau - \epsilon)$. Note that the situation here 
is the reverse of that in real electromagnetism, where ${\bf B}$ changes 
sign under time reversal and ${\bf E}$ is invariant.
\end{itemize}

These symmetries are all present in analogous forms in the diamond lattice 
gauge theory. There is also an additional global symmetry, 
which exchanges the two sublattices of up- and down-pointing sites. 

\section{Ergodicity of Single Ring Moves in a Square Lattice Model}
\label{app:2d-ergodic}

We consider the classical dimer model on the square lattice at infinite temperature,
with two dimers touching every site. This is the $d=2$ analog of the
RK point of our cubic model. We work in the electric field language, where $e_{{\bf r}i} 
= \pm 1/2$
with $i = x,y$, and $(\operatorname{div} e)_{{\bf r}} = 0$.
Consider an $L \times L$ system ($L$ even) with periodic boundary conditions:
$e_{[{\bf r}+L{\bf e}_i],j} = e_{{\bf r}j}$, where ${\bf e}_i = {\bf x},{\bf y}$.
The zero-flux sector is specified by the condition:
\begin{equation}
\sum_{n_x=0}^{L-1} e_{n_x{\bf x},y} = \sum_{n_y=0}^{L-1} e_{n_y{\bf y},x} = 0\text{.}
\end{equation}
We will show that any two states in this sector are connected by a sequence of ring
exchange moves on single square plaquettes.

The key step is to go to a height representation on the dual lattice with sites at the plaquette
centers ${\sf r} = {\bf r} + ({\bf x} + {\bf y})/2$. We define:
\begin{eqnarray}
\label{eq:height-defns}
h({\sf r} + {\bf x}) - h({\sf r}) &=& -2 e_{[{\sf r} + ({\bf x} - {\bf y})/2],y} \\
h({\sf r} + {\bf y}) - h({\sf r}) &=& 2 e_{[{\sf r} + ({\bf y} - {\bf x})/2],x} \nonumber
\end{eqnarray}
The content of this definition is that the height increases/decreases by one if we cross a
link of the direct lattice with electric field pointing to the right/left.
If we fix the value of the
height on one site of the dual lattice these definitions determine it uniquely everywhere. The
height is well-defined because 
the electric field has zero divergence, and can consistently be taken
to have periodic boundary conditions because we are in the zero-flux sector.

A flippable
plaquette has a height either above or below all of its four neighbors, depending on its
orientation. Therefore,
in a given configuration, the plaquettes with minimum and maximum height will always be
flippable, and every configuration has at least two flippable plaquettes. There are also two
states with every plaquette flippable; up to overall shifts of the height these have
\begin{equation}
\label{eq:square-max-flippable}
h\big( n_x{\bf x} + n_y{\bf y} + ({\bf x}+{\bf y})/2\big)
= \pm\frac{1}{2}\big(1 + (-1)^{n_x+n_y}\big)\text{.}
\end{equation}
These two configurations are connected by single ring moves, since we can 
flip all the maximum height plaquettes in one to go to the other.

To complete the proof, we label configurations by $\Delta h = h_{max} - h_{min}$. This clearly
takes on the minimum possible value of unity in only the two maximally flippable states. Suppose
we are in some other state with $\Delta h > 1$. Then we can flip plaquettes of maximum height
until $\Delta h$ is reduced by 1. This procedure can be repeated until $\Delta h = 1$ and we
reach one of the maximally flippable states. We have thus shown that any two states are connected
by a sequence of single ring moves, because the maximally flippable states are connected to all
states and to each other.

\end{document}